\def\sAe{{^{\rm s}\mathfrak{A}}}
\def\sA{{^{\rm s}\boldsymbol{\mathfrak{A}}}}
\def\Ae{\mathfrak{A}}
\def\A{\boldsymbol{\mathfrak{A}}}
\def\De{\mathfrak{D}}
\def\D{\boldsymbol{\mathfrak{D}}}
\def\sD{{^{\rm s}\!D}}
\def\mD{{^{\rm s}\bf D}}

\def\T{{\mathbb T}}
\def\LA{\left\langle}
\def\RA{\right\rangle}
\def\be{\begin{align}}
\def\se#1{\begin{subequations}\label{#1}
\renewcommand{\theequation}{\theparentequation.\arabic{equation}}}
\def\BE{\begin{equation}}
\def\EE#1{\label{#1}\end{equation}}
\def\rf#1{(\ref{#1})}
\def\xf#1{Fig.~\ref{#1}}
\def\I{{\rm i}}
\def\e{{\rm e}}
\def\d{\,{\rm d}}
\def\L{\boldsymbol{\mathfrak{L}}}
\def\M{\boldsymbol{\mathfrak{M}}}
\def\B{\boldsymbol{\mathfrak{B}}}
\def\G{\boldsymbol{\mathfrak{G}}}
\def\gs{\apprge}
\documentclass[12pt,a4paper]{Melsarticle}
\usepackage{amsmath,amssymb,graphicx,wasysym}
\usepackage{lscape}
\begin{document}
\title{Computation of kinematic and magnetic $\alpha$-effect
and eddy diffusivity tensors by Pad\'e approximation}
\author[pof]{S\'\i lvio M.A.~Gama}
\ead{smgama\@fc.up.pt,~http://sigarra.up.pt/fcup}
\author[poe]{Roman Chertovskih}
\author[mitp]{Vladislav Zheligovsky}
\address[pof]{Centro de Matem\'atica da Universidade do Porto,
Faculty of Sciences, University of Porto\\
R.~Campo Alegre 687, 4169-007 Porto, Portugal}
\address[poe]{Research Center for Systems and Technologies, Faculty of
Engineering, University of Porto,\\
Rua Dr.~Roberto Frias, s/n, 4200-465, Porto, Portugal}
\address[mitp]{Institute of Earthquake Prediction Theory and
Mathematical Geophysics, Russian Ac. Sci.,\\
84/32 Profsoyuznaya St, 117997 Moscow, Russian Federation}

\begin{abstract}
We present examples of Pad\'e approximation of the $\alpha$-effect and eddy
viscosity/diffusivity tensors in various flows. Expressions for the tensors
derived in the framework of the standard multiscale formalism are employed.
Algebraically the simplest case is that of a two-dimensional parity-invariant
six-fold rotation-symmetric flow, where eddy viscosity is negative, indicating
intervals of large-scale instability of the flow. Turning to the
kinematic dynamo problem for three-dimensional flows of an incompressible
fluid, we explore application of Pad\'e approximants for computation
of tensors of magnetic $\alpha$-effect and, for parity-invariant
flows, of magnetic eddy diffusivity. We construct Pad\'e approximants
of the tensors expanded in power series in the inverse molecular diffusivity
$1/\eta$ around $1/\eta=0$. This yields the values of the dominant growth rate
due to the action of the $\alpha$-effect or eddy diffusivity to satisfactory
accuracy for $\eta$, several dozen times smaller than the threshold, above which
the power series is convergent. For one sample flow, we observe eddy diffusivity
tending to negative infinity when $\eta$ tends from above to the point
of the onset of small-scale dynamo action in a symmetry-invariant subspace
where a neutral small-scale magnetic mode resides. However, 49 first
coefficients in the power series in $1/\eta$ prove insufficient
for Pad\'e approximants to reproduce this behaviour.
We do computations in Fortran in the standard
``double'' (real*8) and extended ``quadruple'' (real*16) precision, as well as
perform symbolic calculations in Mathematica.
\end{abstract}

\maketitle

\section{Introduction}

Power series expansion of analytic functions is perhaps the most powerful tool
of numerical analysis. Let us just note that most algorithms for numerical
integration of ordinary differential equations, such as the Runge--Kutta
methods, rely on Taylor series expansions for derivation. The truncated
series --- i.e., polynomials --- are easy to compute, and thus provide
an important basic algorithm for evaluation of many analytic functions.

However, there are two caveats. One is related to a finite precision
of computations, stemming from the hardware architecture and employed
in the overwhelming majority of computer codes. A well-known example of the resultant
failure of a computational procedure is a straightforward attempt to compute
by this technique an exponent of a real large negative number \cite{FMM}.
Mathematically, this does not present any difficulty --- the large individual
terms in the Taylor expansion around zero are guaranteed to mutually cancel out
and yield the final result which is less than unity. For finite-precision
computations, however, the cancellation is not any more guaranteed, and
the initial growth of individual terms can result in ultimate loss of accuracy.

The other one stems from finiteness of the radius of convergence of most power
series encountered in computational practice. A complementary technique is then
needed to continue analytically a function defined by the power series outside
its circle of convergence. This can be achieved by constructing the so-called
Pad\'e approximants \cite{Gi,B90,B96}, i.e., an implementation of the continuation
in the form of the ratio of two polynomials. Let us cite the words of appraisal
in \cite{nr}: ``Pad\'e approximation has the uncanny knack of picking the function
you had in mind from among all the possibilities. Except when it doesn't!
That is the downside of Pad\'e approximation: it is uncontrolled. There is,
in general, no way to tell how accurate it is, or how far out in $x$
it can usefully be extended. It is a powerful, but in the end still
mysterious, technique."

A not less mysterious notion is that of eddy diffusivity \cite{St}, also known
as eddy (or turbulent) viscosity when fluid viscosity, the source of diffusion
in hydrodynamics, is considered. Like the magnetic $\alpha$-effect
and anisotropic kinetic alpha- (aka AKA) effect, eddy diffusivities are often
encountered in magnetohydrodynamics when generation of large-scale magnetic
fields by flows of electrically conductive fluids is considered.
At first sight, it is in direct contradiction with the second principle
of thermodynamics. Of course, in fact no physical laws are violated. Both
notions just describe the mean influence of the small scales on large-scale
structures.

According to the modern paradigm, cosmic magnetic fields (such as the solar
or geomagnetic field) exist due to the dynamo processes in the moving
electrically conductive medium (such as melted rocks in the outer Earth's
core) \cite{Mb}. The generating flows are typically turbulent
and feature a vast hierarchy of spatial and temporal scales. Small-scale
fluctuations of the flow (called ``cyclonic events'' by E.~Parker)
give rise to small fluctuations of magnetic field. The interaction
of the small-scale components of the magnetic field and flow velocity
produces a mean electromotive force (e.m.f.) that may have a non-zero component
parallel
to the mean magnetic field, and this can be beneficial for magnetic field
generation \cite{P55}. The part of the mean e.m.f.~linear in the mean field
gives rise to the so-called magnetic $\alpha$-effect. If the flow is
parity-invariant, the $\alpha$-effect disappears and the impact of yet smaller
spatial scales becomes apparent; the mean e.m.f.~is then a linear combination
of the first-order spatial derivatives of the mean field, giving rise to eddy
(turbulent) diffusivity. These physical ideas are treated under
various simplifying assumptions in the mean-field electrodynamics \cite{SKR,KR},
and, relying only on the first principles, by asymptotic methods
of homogenization of elliptic operators in the magnetohydrodynamic multiscale
stability theory \cite{GO1,GO2,V86,V87,DF,La,VZ}. Weakly
nonlinear stability problems are also amenable to these methods~\cite{CZ}.

Analysis of equations makes it evident that eddy viscosity/diffusivity acquires
the unusual properties, because it acts on mean fields only, i.e., essentially
an open physical system is considered. In fact, in this class of MHD systems
the inverse energy cascade is important, energy proliferating from small scales
towards large ones; the source of energy for the developing large-scale magnetic,
hydrodynamic or combined MHD perturbation is the forcing applied to maintain
the perturbed (also sometimes called basic) fluid flow.

Evaluation of the $\alpha$-effect and eddy diffusivity tensors involves
solving the so-called auxiliary problems, which are linear problems
for the respective elliptic operators of linearization. In the large-scale
dynamo problem, computing the magnetic $\alpha$-effect tensor requires
considering three such problems; the number increases to 12, when
the tensor of eddy diffusivity is sought (unless auxiliary problems
for the adjoint operator come into play, decreasing the number of auxiliary
problems to be treated to just 6, see, e.g., \cite{ABNZ,RCZ}). Interesting
results (e.g., instability to large-scale perturbation or dynamos)
are typically obtained for relatively small molecular
viscosity or magnetic diffusivity. Consequently, high spatial resolution
is needed when solving the auxiliary problems, which makes the problems
computationally intensive. However, the tensors can be easily expanded in
the respective Reynolds number (i.e., in the inverse viscosity or diffusivity
provided the size of the flow periodicity box and the flow velocity are order
unity), when it is small, i.e., for large viscosities and diffusivities. When
the parameter tends to the critical value for the onset
of the small-scale instability (i.e., in the dynamo context, to the value
for which generation of small-scale magnetic field starts),
the tensors usually exhibit singular behaviour \cite{ZPF,RCZ,ACZ,ACZ2} of
a simple pole type, bounding from above the radius of convergence of the series.
This suggests to try Pad\'e approximants for computing the tensors
and the respective large-scale magnetic field / instability growth rates.

We report here numerical experiments exploring these ideas.
The paper is organized as follows. In section~\ref{sec:edvis} we apply
the Pad\'e approximants techniques for evaluation of the eddy viscosity
in a two-dimensional flow with two symmetries, in whose presence the eddy
viscosity tensor reduces to a scalar. In view of the first caveat
discussed in the beginning of this introduction, we have chosen to perform
the calculations in precise arithmetics allowing an arbitrary number
of correct digits; for this purpose, we have used the programming language
Mathematica, giving an opportunity to make symbolic computations.
In section~\ref{sec:mae} we revert to the standard ``double precision''
computations (real*8, in Fortran speak) of the magnetic $\alpha$-effect tensor,
using the ``quadruple precision'' (real*16) computations for comparison.
In section~\ref{sec:med} we again employ the symbolic capabilities
of Mathematica to evaluate the magnetic eddy diffusivity tensor.
Our findings are summarized in section~\ref{sec:co}.

\section{Calculation of eddy viscosity}\label{sec:edvis}

No truly two-dimensional flows exist in nature, but they mimick properties of
natural objects, such as the atmosphere or ocean \cite{Tay,Fb,Lin}.
We analyze the eddy viscosity tensor, $\varepsilon_{ijk\ell}$, \cite{DF}
of a two-dimensional flow of incompressible fluid that has two symmetries:
parity invariance ($S1$) and the six-fold rotation symmetry ($S2$).

Since an $S1$-symmetric flow has a center of symmetry, it cannot possess
the large-scale anisotropic kinetic $\alpha$-effect \cite{FSS}.
This is important, because in the presence of the AKA effect
the large-scale dynamics is essentially dispersive and non-diffusive,
concealing the impact of the eddy viscosity. The symmetry $S2$ implies
the isotropy of fourth-order tensors (see, for instance, \cite{LL}),
in particular, $\varepsilon_{ijk\ell}=\nu_{\rm E}\delta_{ij}\delta_{k\ell}$,
where the scalar $\nu_{\rm E}$ is the (standard) eddy viscosity and
$\delta_{mn}$ is the Kronecker symbol. Although the assumption that a flow
features the two symmetries is mathematically convenient, it may be
non-realistic when considering natural or engineering problems~\cite{CH}.

Two-dimensional flows endowed with the symmetries $S1$ and $S2$ can be
constructed as follows. A~space-periodic flow is assumed, the periodicity cell
being the rectangle
$$[0,L_1]\times[0,L_2]\ni{\bf x}=(x_1,x_2),\qquad L_1=\sqrt{3}L_2=2\pi.$$
Its stream-function $\Psi(t,{\bf x})$ is then a sum of Fourier modes, whose
wave vectors are $p(2,0)+q(1,\sqrt{3})$, where $p$ and $q$ are integer.
Any two such modes, that have wave vectors mutually related by rotations
by $\pi/3$, must both be involved in the sum with the same real amplitude.

We begin this section by recalling the expression for the scalar eddy viscosity
in terms of the solutions to two auxiliary problems and the analytical
framework for evaluating the expansion of the eddy viscosity tensor in powers
of the inverse of the molecular viscosity. We carry on by recalling
the standard terminology and definitions of Pad\'e
approximants to a series. Finally, we discuss our results and conclusions.

\subsection{Eddy viscosities and multiscale techniques}

For parity-invariant and six-fold rotation-symmetric flows of incompressible
fluid, eddy viscosities were calculated in \cite{DF} by multiscale techniques.
They can be expressed in terms of solutions to two auxiliary problems,
which can be solved analytically only in special cases \cite{DF}
(e.g., if the flow depends only on a single spatial coordinate).

Briefly, the eddy viscosity in an isotropically forced
two-dimensional flow is calculated as follows \cite{DF,GVF}. In terms
of the stream-function $\Psi(t,{\bf x})$, the two-dimensional forced
Navier--Stokes equation for incompressible fluid takes the form
$$\partial_t\nabla^2\Psi+J(\nabla^2\Psi,\Psi)=
\nu\nabla^2\nabla^2\Psi+\partial f_1/\partial x_2-\partial f_2/\partial x_1.$$
Its solution, $\Psi$, defines a {\it basic flow}. Here,
$J(g_1,g_2)=(\partial g_1/\partial x_1)(\partial g_2/\partial x_2)
-(\partial g_1/\partial x_2)(\partial g_2/\partial x_1)$
is the Jacobian determinant of functions $g_1(x_1,x_2)$ and $g_2(x_1,x_2)$,
the operator $\nabla^2=\partial^2\!/\partial x^2_1+\partial^2\!/\partial x^2_2$
is the Laplacian,
$\nu$ the kinematic molecular viscosity, and ${\bf f}=(f_1,f_2)$ the external
force. (In our numerical examples, the flow norm and the size
of the small-scale periodicity cell are order unity; thus, the inverse
molecular viscosity can be regarded as the local Reynolds
number, which is the key dimensionless parameter of the problem.)
Now assume that the basic flow possesses the symmetries $S1$ and $S2$. Then
the (scalar) eddy viscosity coefficient, $\nu_{\rm E}=\nu_{\rm E}(\Psi,\nu)$,
that depends only on the basic flow and molecular viscosity, is \cite{GVF}
\BE\nu_{\rm E}(\Psi,\nu)=\nu-\LA\left(Q+2\,{\partial S\over\partial x_1}\right)
{\partial\Psi\over\partial x_2}\RA.\EE{eq:nueddy}
Here, the angle brackets denote the average over the periodicity cell:
$$\LA g(x_1,x_2)\RA=\frac{1}{L_1L_2}\int_{x_1=0}^{L_1}\int_{x_2=0}^{L_2}
g(x_1,x_2)\d x_2 \d x_1,$$
the scalar functions $Q(x_1,x_2)$ and $S(x_1,x_2)$ are
solutions to two auxiliary problems
\BE\G Q={\partial\nabla^2\Psi\over\partial x_2},\qquad
\G S=Q\,{\partial\nabla^2\Psi\over\partial x_2}
+2J\left(\Psi,{\partial Q\over\partial x_1}\right)
-\nabla^2Q\,{\partial\Psi\over\partial x_2}
+4\nu\,{\partial\nabla^2 Q\over\partial x_1},\EE{eq:aux1}
and $\G$ denotes the linearization of the Navier--Stokes equation
around $\Psi$,
$$\G:\psi\mapsto J(\nabla^2\psi,\Psi)+J(\nabla^2\Psi,\psi)
-\nu\nabla^2\nabla^2\psi.$$
We restrict it to zero-mean functions of the same space periodicity
as the basic flow. In this functional space we can define the inverse
Laplace operator, that we denote $\nabla^{-2}$. A field $G$ from this space
can be expressed as a Fourier series
\BE G(x_1,x_2)=\sum_{p,q\in\mathbb{Z}}\widehat{G}_{p,q}
\exp(2\pi\I(px_1/L_1+qx_2/L_2)),\EE{eq:exp1}
where $\LA G\RA=\widehat{G}_{0,0}=0$. The equation
$\nabla^2F=G\ \Leftrightarrow\ F=\nabla^{-2}G$ can now be readily solved:
given $\LA F\RA=0$, we find
\BE\nabla^{-2}G=-(2\pi)^{-2}\sum_{p,q\in\mathbb{Z}\backslash\{0\}}
{\widehat{G}_{p,q}\over(p/L_1)^2+(q/L_2)^2}
\,\exp(2\pi\I(px_1/L_1+qx_2/L_2)).\EE{eq:exp2}

Existence of a deterministic time-independent space-periodic flow, which
has an isotropic negative eddy viscosity when the molecular viscosity is below
a critical value, was established in \cite{VGF}. The so-called
decorated hexagonal flow (DHF), on which we will also focus here, is
\BE\begin{array}{rl}
\Psi(x_1,x_2)=\big[\!\!\!
&-\cos2x_1-\cos(x_1+\sqrt{3}x_2)-\cos(x_1-\sqrt{3}x_2)\\
&+\cos(4x_1+2\sqrt{3}x_2)+\cos(5x_1-\sqrt{3}x_2)+\cos(x_1-3\sqrt{3}x_2)\\
&-\cos4x_1-\cos(2x_1+2\sqrt{3}x_2)-\cos(2x_1-2\sqrt{3}x_2)\\
&+\cos(4x_1-2\sqrt{3}x_2)+\cos(5x_1+\sqrt{3}x_2)+\cos(x_1+3\sqrt{3}x_2)
\big]/2.\end{array}\EE{eq:natoflow-1}

The phenomenon of negative eddy viscosity is quite common among
two-dimensional divergenceless space-periodic basic flows
with the symmetries $S1$ and $S2$: about 1/3 of such flows feature
negative eddy viscosity for sufficiently low molecular viscosity \cite{GVF}.
The auxiliary problems \rf{eq:aux1} can be solved either numerically
by spectral methods, or by expanding in powers of $\nu^{-1}$ to high orders
and afterwards extending analytically (relying on their meromorphy
\cite{GVF}) beyond the disk of convergence. We examine the latter method
enabling us to perform all calculations exactly.

\subsection{Eddy viscosity expansion in powers of $\nu^{-1}$}

Let us expand \rf{eq:nueddy}:
\BE\nu_{\rm E}(\Psi,\nu)=\nu+\sum_{n=1}^\infty\nu_{\rm E}^{(n)}(\Psi)\nu^{-n}.
\EE{eq:nu_E_exp}
To calculate $\nu_{\rm E}^{(n)}$, we expand the solutions $Q$ and $S$
to \rf{eq:aux1} in Maclaurin series in $\nu^{-1}$:
$$Q=\sum_{n=1}^\infty Q^{(n)}\nu^{-n},\qquad S=\sum_{n=1}^\infty S^{(n)}\nu^{-n}.$$
Substituting the series into \rf{eq:nueddy} and integrating by parts yields
\BE\nu_{\rm E}^{(n)}(\Psi)=\LA-Q^{(n)}{\partial\Psi\over\partial x_2}
+2S^{(n)}{\partial^2\Psi\over\partial x_1\partial x_2}\RA.\EE{eq:nu_Ei}
Here $Q^{(1)}=-\nabla^{-2}(\partial\Psi/\partial x_2)$ and
$S^{(1)}=-4\nabla^{-2}(\partial Q^{(1)}/\partial x_1)$,
and the subsequent terms satisfy the recurrence relations
\se{rec}\hspace*{-2em}\be Q^{(n)}&=\B Q^{(n-1)},\\
S^{(n)}&=\B S^{(n-1)}-\!\nabla^{-2}\nabla^{-2}\left[
Q^{(n-1)}{\partial\nabla^2\Psi\over\partial x_2}
+2J\left(\!\Psi,{\partial Q^{(n-1)}\over\partial x_1}\right)
-\!\nabla^2Q^{(n-1)}\,{\partial\Psi\over\partial x_2}
+4{\partial\nabla^2Q^{(n)}\over\partial x_1}\right]\!,
\end{align}\end{subequations}
where the operator $\B$ is defined as
$$\B:f\mapsto\nabla^{-2}\nabla^{-2}\left[J(\nabla^2f,\Psi)
+J(\nabla^2\Psi,f)\right].$$
Since we consider here parity-invariant flows, their stream-functions being even,
i.e., $\Psi(-{\bf x})=\Psi({\bf x})$, the series \rf{eq:nu_E_exp} involves only
odd powers of $\nu^{-1}$ \cite{GVF}, i.e., $\nu_{\rm E}^{(n)}=0$ for all even $n$.

By definition, the $[L/M]$ Pad\'e approximant to a series, whose first
$m\ge L+M+1$ terms are known, is the ratio of a polynomial of degree $\le L$
to a polynomial of degree $\le M$, such that the first $L+M+1$ terms of the
expansion of the ratio coincide with the respective terms of the series.

We use Pad\'e approximants to reconstruct the dependence of the eddy viscosity
on the inverse molecular viscosity employing the expansion \rf{eq:nu_E_exp}.
The Pad\'e approximation techniques can also be naturally applied for exploring
the poles of the eddy viscosity. A pole on the real axis can usually be linked
to the onset of linear instability to small-scale perturbations, or sometimes
(if it appers again on decreasing $\nu>0$) to its cessation.

\subsection{Results of calculations}

The performance of present-day computers and efficiency of symbolic programming
software gives an opportunity to calculate the coefficients \rf{eq:nu_Ei}
using recurrence relations \rf{rec}, and construct the Pad\'e approximants
exactly. All calculations reported
in this section are performed with full precision by Mathematica \cite{HMM}.
Table \ref{tab:Prime-decomp} shows some of the first twenty non-zero
coefficients $\nu_{\rm E}^{(n)}$ ($n=1,3,5,...,39$) for the DHF.
One of the reasons to perform full precision symbolic computations
by Mathematica has been a hope to discover useful relations between
the coefficients of the approximants. Unfortunately, none are visible
in the data of Table~\ref{tab:Prime-decomp}.


We truncate the series \rf{eq:nu_E_exp}
at orders up to 39, even terms missing. Roots of their Pad\'e
approximants quickly stabilize near $\nu=\nu^\star\approx0.58$ (see
\xf{pade}), indicating a transition to negative eddy viscosity at lower
molecular viscosities. The root $\nu^\star$ is simple;
a sharpened estimate is $1/\nu^\star=1.72144\pm10^{-5}$.


\begin{landscape}\begin{table}
\renewcommand{\arraystretch}{1.5}
\caption{First 39 non-zero coefficients
of \rf{eq:nu_E_exp} for the DHF (prime decomposition, where presented).
The first 5 significant figures of these coefficients are given in \cite{VGF}.
$``a\lll p\ggg c"$ denotes a natural number containing $p$ decimal
digits between $a$ and $b$.}

\begin{center}\begin{tabular}{c|c}
$n$ & Coefficient $\nu_{\rm E}^{(n)}$ (exact rational number)\\\hline
1 & $\frac{3}{2^2}$\\\hline
3 & $-\frac{3\times5\times11\times1931\times80491}{2^9\times7^4\times13^2\times19^2}$\\\hline
5 & $-\frac{3\times53^2\times222967\times1994517983033813651288306079222192539}{2^{19}\times5^2\times7^9\times13^6\times19^7\times31^3\times37^2\times43^3\times61^2}$\\\hline
7 & $\frac{3^3\times23\times17401\times11608063\times570396658307516795186040829874710499\times595146062519802577066082838776447096016784218965582671080441286999}{2^{25}\times5^2\times7^{17}\times13^{10}\times19^{11}\times31^5\times37^7\times43^5\times61^6\times67^3\times73^3\times79^3\times97^3\times103^2\times109^2}$\\\hline
9 & $-\frac{9606359879\:\lll188\ggg\:5777697637}{2^{33}\times5^2\times7^{24}\times11^2\times13^{14}\times19^{15}\times31^7\times37^{11}\times43^7\times61^{10}\times67^5\times73^5\times79^5\times97^5\times103^3\times109^6\times127^3\times139^3\times151^3\times157^2\times163^2}$\\\hline
11 & $-\frac{7129561983\:\lll324\ggg\:7108258721}{8493879641\:\lll326\ggg\:4312960000}$\\\hline
$\vdots$ & $\vdots$\\\hline
39 & $-\frac{1648936106\:\lll9785\ggg\:2091564067}{3775138782\:\lll9788\ggg\:0000000000}$\\\hline
\end{tabular}\end{center}\label{tab:Prime-decomp}\end{table}

\begin{figure}
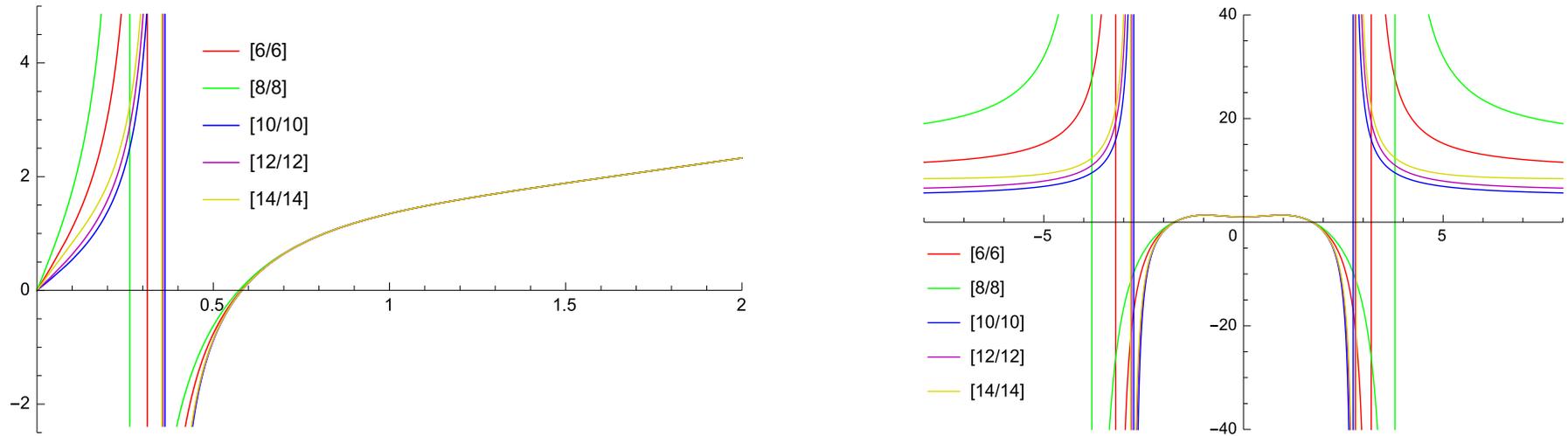

\centerline{\includegraphics[height=65mm]{pad6_14a.ps}\hspace*{6em}
\includegraphics[height=65mm]{pad6_14b.ps}}
\caption{Pad\'e approximants (vertical axes) $[L/L]_{\nu_{\rm E}}(\nu)$ (left)
and $[L/L]_{\nu_{\rm E}/\nu}(\nu^{-1})$ (right) for $6\le L\le14$ step 2.
Horizontal axes: $\nu$ (left), $\nu^{-1}$ (right).
In the right panel, we extend on purpose $\nu$
to negative values to highlight that the Maclaurin expansion of
$\nu_{\rm E}\left(\nu^{-1}\right)/\nu$ is an even function of $\nu^{-1}$.}
\label{pade}\end{figure}\end{landscape}

%
%
%

These results indicate that Pad\'e approximants is a reliable alternative
to other methods for calculation of the point of the onset of large-scale
instability in this type of flows. Not only they furnish stable
estimates (provided the approximated function is meromorphic and the series is
long enough), but also serve as precursors to future work. An illustration is
\xf{pade} (right), where the $[14/14]$ approximant (as many others) exhibits
a singular behaviour near $1/\nu\approx2.81$ apparently related to the onset
of linear instability to small-scale perturbations. The non-monotonicity
of the eddy viscosity as a function of $\nu$ can be regarded as a manifestation
of the complexity of the two-dimensional turbulent flow.

\section{Computation of the magnetic $\alpha$-effect tensor}\label{sec:mae}

As we have seen in the previous section, the use of Pad\'e expansions
for reconstructing the dependence of eddy viscosity on the molecular one
is possible and does not require very high degrees of the polynomials involved
at least for moderate (not very small) molecular viscosities. However,
arbitrary-precision symbolic calculations cannot be regarded as a practical
realization of the approximation algorithm. Here we construct
Pad\'e approximants for evaluation of the magnetic $\alpha$-effect tensor
using arithmetics of floating point numbers of the conventional double (real*8)
and the extended quadruple (real*16) precision. The problem now at hand is
to construct approximations applicable for small magnetic molecular diffusivities.

\subsection{The multiscale formalism revealing the magnetic $\alpha$-effect}

We review here the multiscale expansions \cite{VZ} describing the kinematic
generation of large-scale magnetic field by small-scale zero-mean
space-periodic steady flows. Our magnetic modes depend on two three-dimensional
spatial variables, the fast, $\bf x$, and slow, ${\bf X}=\varepsilon\bf x$, one
(the flow $\bf v$ depends exclusively on~$\bf x$). A magnetic
mode $\bf b$ is an eigenfield of the magnetic induction operator $\L$:
\se{oL}\be\L{\bf b}&=\lambda{\bf b},\label{Leig}\\
\L:{\bf b}&\mapsto\eta\nabla^2{\bf b}+\nabla\times({\bf v}\times{\bf b}).
\label{Lop}\end{align}\end{subequations}
Here $\eta$ is the magnetic molecular diffusivity and Re\,$\lambda$ the growth
rate of the mode $\bf b$. Both the mode and flow are solenoidal.

The scale ratio $\varepsilon$ is a small parameter of the problem, in which
the magnetic mode $\bf b$ and its growth rate are expanded:
\se{ble}\be{\bf b}&=\sum_{n=0}^\infty{\bf b}_n({\bf X},{\bf x})\,\varepsilon^n,\label{bex}\\
\lambda&=\sum_{n=0}^\infty\lambda_n\varepsilon^n.\label{lex}\end{align}\end{subequations}
By substituting the expansions into \rf{Leig} and the solenoidality conditions,
we derive a hierarchy of equations for the coefficients in \rf{ble}.

Like in the previous section, we denote by angle brackets the mean over
the periodicity cell $\T^3$ in the fast variables and by the braces
the fluctuating part:
$$\LA{\bf f}\RA=(2\pi)^{-3}\int_{\T^3}{\bf f}({\bf X},{\bf x})\,\d{\bf x}
=\sum_{k=1}^3\LA{\bf f}\RA_k{\bf e}_k,\qquad\bf\{f\}=f-\LA f\RA.$$
Here ${\bf e}_k$ are unit Cartesian coordinate vectors.

The relevant solution to the first (order $\varepsilon^0$) equation
$\L{\bf b}_0=\lambda_0{\bf b}_0$ is $\lambda_0=0$ and a linear combination
${\bf b}_0=\sum_{k=1}^3\LA{\bf b}_0\RA_k{\bf s}_k$
of small-scale solenoidal neutral magnetic modes ${\bf s}_k({\bf x})$ that are
solutions to the three so-called {\it auxiliary problems of type~I}:
\BE\L{\bf s}_k=0,\qquad\LA{\bf s}_k\RA={\bf e}_k.\EE{Seq}

Averaging the second (order $\varepsilon^1$) equation,
\BE\L{\bf b}_1+2\eta(\nabla\cdot\nabla_{\bf X}){\bf b}_0
+\nabla_{\bf X}\times({\bf v}\times{\bf b}_0)=\lambda_1{\bf b}_0,\EE{eq1}
we obtain an eigenvalue problem
\BE\nabla_{\bf X}\times(\A\LA{\bf b}_0\RA)=\lambda_1\LA{\bf b}_0\RA,\qquad
\nabla_{\bf X}\cdot\LA{\bf b}_0\RA=0\EE{aleq}
(the subscript $\bf X$ denotes differentiation in the slow variables).
Here $\A$ is the tensor of magnetic $\alpha$-effect. The $k$th column
of this $3\times3$ matrix is
\BE\A_k=\LA{\bf v}\times{\bf s}_k\RA,\EE{Adef}
in agreement with the Parker's \cite{P55} concept of the interaction
of fine structures of the flow, $\bf v$, and magnetic field,
$\sum_{k=1}^3\LA{\bf b}_0\RA_k\{{\bf s}_k\}$, giving rise
to a mean e.m.f., $\A\LA{\bf b}_0\RA$, linear in the large-scale
magnetic field $\LA{\bf b}_0\RA$. For space-periodic mean fields
\BE\LA{\bf b}_0\RA={\bf B}\e^{\I\bf q\cdot X},\qquad{\bf B}\cdot{\bf q}=0\EE{mm}
where $\bf q$ is an arbitrary unit vector, straightforward algebra
\cite{RCZ} yields solutions to the eigenvalue problem \rf{aleq}:
\BE\lambda_{1_\pm}({\bf q})={\I\over2}\left((\Ae^2_3-\Ae^3_2)q_1
+(\Ae^3_1-\Ae^1_3)q_2+(\Ae^1_2-\Ae^2_1)q_3\right)\pm\sqrt a,\qquad
a={\bf q}\cdot(\det\sA)\,\sA^{-1}{\bf q}.\EE{ei}
Here $\sAe_k^l=(\Ae_k^l+\Ae_l^k)/2$
are entries of the symmetrized $\alpha$-tensor $\sA=(\A+\A^*)/2$.

For $a\le0$, the $\alpha$-effect just sustains harmonic oscillations
of the mean magnetic field in the slow time $T_1=\varepsilon t$. When $a>0$,
the slow-time growth rate Re$\,\lambda_1({\bf q})=\sqrt a$ of the large-scale
magnetic mode depends only on the symmetrized tensor $\sA$, whose
eigenvalues $\alpha_i$ are real. In the Cartesian coordinate
system, whose axes coincide with eigenvectors of $\sA$, \rf{ei} takes the form
$$a=\alpha_1\alpha_2(q'_3)^2+\alpha_2\alpha_3(q'_1)^2
+\alpha_1\alpha_3(q'_2)^2,$$
where $q'_i$ are components of $\bf q$ in this basis. Thus,
\BE\gamma_\alpha\equiv\max_{|{\bf q}|=1}{\rm Re}\,\lambda_{1_\pm}({\bf q})=
\sqrt{\max(\alpha_1\alpha_2,\alpha_2\alpha_3,\alpha_1\alpha_3)}\EE{mgr}
is the maximum slow-time growth rate of large-scale magnetic modes generated
by the $\alpha$-effect. While the entries of the $\alpha$-effect tensor,
$\Ae^k_l$, are smooth functions of $\eta$, the graph of $\gamma_\alpha$ \rf{mgr}
has cusps at the points $\eta$, where $\alpha_1<\alpha_2=0<\alpha_3$ \cite{RCZ}
(see, e.g., two such cusps in \xf{trf}).

When $a\ne0$ and the kernel of the magnetic induction operator $\L$ does not
involve small-scale zero-mean modes (generically both conditions
are satisfied), all terms in the expansions \rf{ble} can be determined
from the hierarchy of equations obtained by substituting the series
into the eigenvalue equation \rf{Leig}. If the symmetrized tensor $\sA$
is positively or negatively defined (and if the spatial periodicity
of the eigenfunction is compatible with that of the flow), then the series
\rf{ble} are summable \cite{V87} for sufficiently small $\varepsilon$
and constitute an analytical in $\varepsilon$
eigensolution for the large-scale magnetic induction operator; a unique
$\varepsilon$-parameterized branch of the eigenvalues \rf{lex} originates
from any simple eigenvalue $\lambda_1$ of the $\alpha$-effect operator.

\subsection{Pad\'e approximation}\label{pa}

The modes ${\bf s}_k$ (and, consequently, elements of the magnetic
$\alpha$-effect tensor) are functions of molecular eddy diffusivity $\eta$,
meromorphic in this parameter. (By contrast, the slow-time growth rates
of modes generated by the $\alpha$-effect are not, because the square root
present in \rf{ei} gives rise to branch points.)
A power series expansion of ${\bf s}_k$ in $\eta^{-1}$ for large $\eta$
can be constructed like in the hydrodynamic problem considered
in the previous section. We divide \rf{Seq} by $\eta$ to obtain
$$\nabla^2{\bf s}_k=-\eta^{-1}\nabla\times({\bf v}\times{\bf s}_k).$$
Consequently, the coefficients in the expansion
\BE{\bf s}_k=\sum^\infty_{n=0}{\bf s}_k^{(n)}\eta^{-n}\EE{Se}
satisfy the recurrence relations
\BE{\bf s}^{(0)}_k={\bf e}_k,\qquad{\bf s}^{(n)}_k=-\nabla^{-2}\left(\nabla
\times({\bf v}\times{\bf s}^{(n-1)}_k)\right)\quad\mbox{for~}n\ge1\EE{rrS}
(cf.~(5.14)--(5.15) in \cite{GO1}).
Here $\nabla^{-2}$ denotes the inverse Laplacian in the fast variables
acting in the functional space of zero-mean fields.
(Actually, we consider the problem for a flow, whose r.m.s.~velocity is unity;
the size of the periodicity box also being order unity, the inverse molecular
diffusivity $\eta^{-1}$ can be regarded, like in the hydrodynamic case, to be
equal to the local magnetic Reynolds number, the dimensionless parameter
characterizing the mathematical properties of the problem; \rf{Se} can
thus be understood as an expansion in a small Reynolds number.)
These recurrence relations can be used to compute the coefficients
in \rf{Se} by pseudospectral methods. It must be noted that mathematically
they are perfectly suitable for numerical work: indeed, the presence
of the inverse Laplacian in the second relation \rf{rrS} suggests, that
on increasing the number $n$ of the coefficient ${\bf s}^{(n)}_k$ they become
smoother and their energy spectrum decay is steeper. This is in sharp contrast,
for instance, with the recurrence relations for the Lagrangian time-Taylor
coefficients in the expansions of solutions to the Euler equation
for incompressible fluid flow \cite{PZF}.

It is straightforward to determine the radius of convergence of the series
\rf{Se}, $\rho$, regarded as a function
of $1/\eta$. The recurrence relations \rf{rrS} involve the operator
\BE\M:{\bf s}\mapsto-\nabla^{-2}\left(\nabla\times({\bf v}\times{\bf s})\right)\EE{rr}
acting in the functional space of solenoidal zero-mean space-periodic fields.
Since it is compact, its spectrum consists of a countable set of eigenvalues
$\mu_i$, tending to zero. Consequently, \hbox{$\rho\ge 1/\max_i|\mu_i|$};
generically the equality holds, but $\rho>1/\max_i|\mu_i|$, if the expansion
of ${\bf s}^{(1)}_k$ in the basis of eigenfunctions of the operator $\M$
does not involve eigenfunctions associated with any eigenvalue $\mu_i$ such
that $|\mu_i|=\max_i|\mu_i|$. Therefore, generically the radii
of convergence of the series for all the three ${\bf s}_k$ are the same.

Clearly, the radius of convergence of the ensuing series
for the $\alpha$-effect tensor \rf{Adef},
\BE\A_k=\sum^\infty_{n=1}\A_k^{(n)}\eta^{-n},\qquad
\A_k^{(n)}=\LA{\bf v}\times{\bf s}_k^{(n)}\RA\EE{ate}
is not smaller than that of the series \rf{Se} for ${\bf s}_k$. Let us note
a symmetry property of \rf{ate}. Denote by the superscript minus objects
pertinent to the reverse flow $-\bf v$:
\BE\L^-:{\bf b}\mapsto\eta\nabla^2{\bf b}-\nabla\times({\bf v}\times{\bf b}),
\qquad\L^-{\bf s}^-_k=0,\quad\LA{\bf s}^-_k\RA={\bf e}_k,
\qquad\A_k^-=\LA-{\bf v}\times{\bf s}^-_k\RA.\EE{iv}
The $\alpha$-effect tensor $\A^-$ for the reverse flow $-\bf v$ is obtained
from the tensor $\A$ for $\bf v$ by transposition, i.e. ${(\A^-_k)}_l=(\A_l)_k$
\cite{RCZ}. Since the recurrence relations \rf{rrS} are linear in $\bf v$,
\BE{\bf s}^{(n)}_k=(-1)^n({\bf s}^-_k)^{(n)},\EE{rmv}
where $({\bf s}^-_k)^{(n)}$ denote the coefficients in the expansion of
${\bf s}^-_k$ in power series \rf{Se}. This implies
$${(\A_l)}^{(n)}_k=(\A^-_k)^{(n)}_l=\LA-{\bf v}\times({\bf s}^-_k)^{(n)}\RA_l
=(-1)^{n+1}\LA{\bf v}\times{\bf s}_k^{(n)}\RA_{\!l}=(-1)^{n+1}{(\A_k)}^{(n)}_l.$$
Therefore, the coefficients in the series \rf{ate} are symmetric matrices
for odd $n$, and antisymmetric ones for even~$n$. In other words,
the symmetrized $\alpha$-effect tensor $\sA$ involved in computation of
the discriminant~$a$ in \rf{ei} is expanded in odd powers of $\eta^{-1}$,
and of the common imaginary part of $\lambda_{1_\pm}({\bf q})$ in even powers;
the latter expansion is not needed when computing just the growth rates.

\subsection{Numerical results}

Here, we construct Pad\'e approximants for the $\alpha$-effect tensor
\rf{ate} for a sample solenoidal flow, and compare the maximum growth rate
values $\gamma_\alpha$ \rf{mgr} obtained for the approximated tensor to those
computed directly at individual values of $\eta$ by spectral methods.

For this purpose, a sample solenoidal flow has been synthesized as a Fourier
series with pseudo-random coefficients, corrected to make it solenoidal and
zero-mean. It involves Fourier harmonics for wave numbers not exceeding 10.
The coefficients are scaled so that the energy spectrum decays exponentially
by 10 orders of magnitude and the r.m.s.~velocity is unity.

Solutions to auxiliary problems have been computed by the code \cite{Zh}
employing standard pseudo-spectral methods. For $\eta>0.05$,
the problem was preconditioned by the operator $(-\nabla^2)^{-1/2}$,
readily available in the Fourier space. The resolution of $64^3$ Fourier
harmonics was used. Energy spectra of the neutral modes ${\bf s}_k$ decay
for this flow by at least 9 orders of magnitude for the smallest considered
$\eta=0.035$. The Lebesgue space $L_2$ norms of ${\bf s}^{(n)}_k$ from $n=0$
to $n=128$ decay by 38 orders of magnitude, and hence the power series \rf{Se}
and \rf{ate} converge for $\eta\gs0.50$ .

\subsubsection{Approximation by the algorithm \cite{GGT}}

We have tried two approaches for construction of Pad\'e
approximants of the entries of the $\alpha$-effect tensor.
Here we discuss the results obtained by the algorithm proposed in \cite{GGT}.

Pad\'e approximant $[M/L]_f$ of a function
$f(y)=\sum_{n=0}^\infty f^{(n)}y^n$ is the ratio of two polynomials
of degrees $M$ (numerator) and $L$ (denominator), whose $M+L+1$ first Taylor
expansion coefficients coincide with those of $f$. For the sake of argument,
let us assume $M\ge L$. Then $L+1$ coefficients of the denominator
$d(y)=\sum_{n=0}^{L+1}d_ny^n$ satisfy the linear system of equations of the form
\BE\left[\begin{array}{llcll}
f^{(M+1)}&f^{(M)}&...&f^{(M+2-L)}&f^{(M+1-L)}\\
f^{(M+2)}&f^{(M+1)}&...&f^{(M+3-L)}&f^{(M+2-L)}\\
&&...&&\\
f^{(M+L-1)}&f^{(M+L-2)}&...&f^{(M+2)}&f^{(M+1)}\\
f^{(M+L)}&f^{(M+L-1)}&...&f^{(M+1)}&f^{(M)}\end{array}\right]
\left[\begin{array}{l}d_0\\d_1\\...\\d_{L-1}\\d_L\end{array}\right]=0\EE{Toe}
(finding the coefficients of the numerator upon solving \rf{Toe}
is straightforward, see \cite{Gi,B96} for details).

\begin{figure}[t]
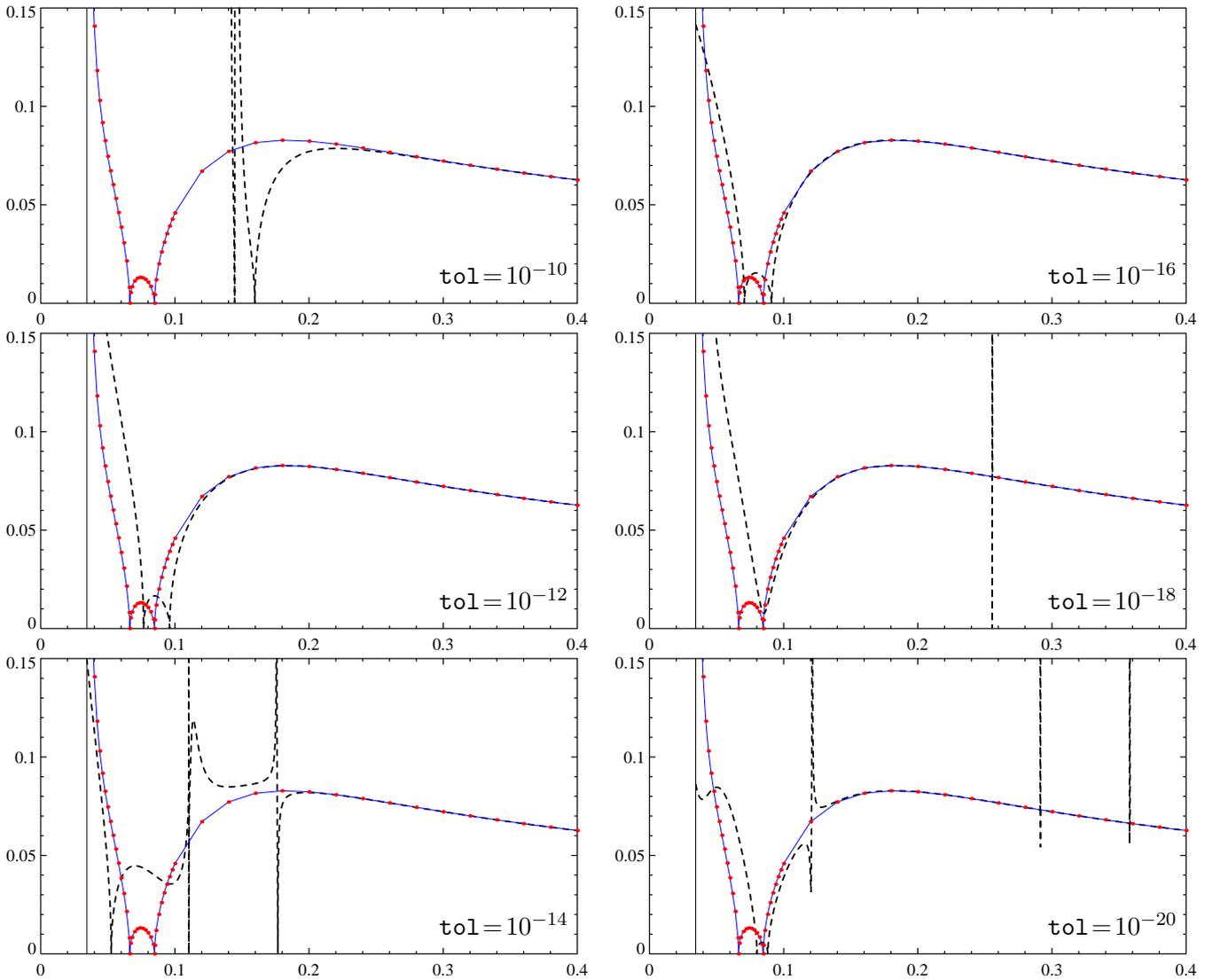

\begin{minipage}[t]{.49\textwidth}
\includegraphics[width=\textwidth]{tr1e-10.ps}
\hspace*{-.32\textwidth}\raisebox{.25in}{\makebox[.3\textwidth]{
\small${\tt tol}\!=\!10^{-10}$}}
\end{minipage}\hfill\begin{minipage}[t]{.49\textwidth}
\includegraphics[width=\textwidth]{tr1e-16.ps}
\hspace*{-.32\textwidth}\raisebox{.25in}{\makebox[.3\textwidth]{
\small${\tt tol}\!=\!10^{-16}$}}
\end{minipage}

\begin{minipage}[t]{.49\textwidth}
\includegraphics[width=\textwidth]{tr1e-12.ps}
\hspace*{-.32\textwidth}\raisebox{.25in}{\makebox[.3\textwidth]{
\small${\tt tol}\!=\!10^{-12}$}}
\end{minipage}\hfill\begin{minipage}[t]{.49\textwidth}
\includegraphics[width=\textwidth]{tr1e-18.ps}
\hspace*{-.32\textwidth}\raisebox{.25in}{\makebox[.3\textwidth]{
\small${\tt tol}\!=\!10^{-18}$}}
\end{minipage}

\begin{minipage}[t]{.49\textwidth}
\includegraphics[width=\textwidth]{tr1e-14.ps}
\hspace*{-.32\textwidth}\raisebox{.25in}{\makebox[.3\textwidth]{
\small${\tt tol}\!=\!10^{-14}$}}
\end{minipage}\hfill\begin{minipage}[t]{.49\textwidth}
\includegraphics[width=\textwidth]{tr1e-20.ps}
\hspace*{-.32\textwidth}\raisebox{.25in}{\makebox[.3\textwidth]{
\small${\tt tol}\!=\!10^{-20}$}}
\end{minipage}

\caption{Maximum slow-time growth rates $\gamma_\alpha$ \rf{mgr} (vertical axis)
of large-scale magnetic modes generated by the $\alpha$-effect as a function
of the molecular diffusivity $\eta$ (horizontal axis), computed using
the $\alpha$-effect tensor, Pad\'e-approximated by the algorithm \cite{GGT}
for a varying tolerance {\tt tol}. Thin solid line: the dependence determined
by computation of $\gamma_\alpha$ at individual $\eta$ values (red solid
circles) by spectral methods (resolution $64^3$ Fourier harmonics),
thick dashed line: the approximate dependence.}
\label{trf}\end{figure}

In our case, the coefficients ${\bf s}^{(n)}$ are obtained by applying
iteratively the operator $\M$ \rf{rr}. It is compact, and its eigenvalues,
except for a finite number of them, are below unity in absolute value.
The respective spectral components decay during the iterations according
to the power law and sooner or later reduce in magnitude below the accuracy
of computations. Consequently, for large $L$ the Toeplitz matrix of size $L\times(L+1)$
in the l.h.s.~of \rf{Toe} becomes numerically degenerate, i.e., its rank
effectively falls below $L$. To construct an approximant under
such adverse numerical conditions, it was proposed in \cite{GGT} to compute
the singular value decomposition of this matrix and to regard its effective
rank as equal to the number of singular values whose absolute value exceeds
the given relative tolerance {\tt tol} (i.e., is not smaller than
${\tt tol}\|(f^{(1)},f^{(2)},...,f^{(M+L-1)},f^{(M+L)})\|$, where $\|\cdot\|$
is the standard Lebesgue space
$L_2$ norm), decreasing the degrees of the polynomials involved in the Pad\'e
approximation, $M$ and $L$, by the number of the ``missing'' dimensions.
To counter noise due to rounding errors, ${\tt tol}=10^{-14}$ was
often used in \cite{GGT}.

Beyond poor spectral properties of the system of equations for the Pad\'e
coefficients, there exist two other reasons for amplification
of the numerical noise originally due to round-off errors:\\
$\bullet$ Pseudospectral methods used in computation of space-periodic
solutions ${\bf s}_k$ to the auxiliary problems \rf{Seq} and their coefficients
${\bf s}^{(n)}_k$ \rf{rrS} involve fast Fourier
transforms. These algorithms are very efficient. However, they operate
by computing various linear combinations of the Fourier coefficients.
Typically, at least for moderate molecular diffusivities, the energy spectra
of these fields decay fast. In a sum of a large coefficient
with a small (in absolute values) one, a significant part of the accuracy
of the smaller coefficient is lost.\\
$\bullet$ Insufficiency of the spatial resolution can result in significant
numerical errors. We may note that while increasing the resolution improves
solutions, it aggravates the FFT accuracy problems.

We have tried the algorithm \cite{GGT}
for a set of $\tt tol$ values ranging between $10^{-10}$ and $10^{-20}$
using the MATLAB procedure provided by the authors of \cite{GGT}. We have
computed 65 first coefficients of the power series expansions of
the symmetrized $\alpha$-effect tensor entries $(\sA_k)^{(n)}_l$ (which involve
only odd powers of $1/\eta$, see section \ref{pa}) up to order
$\eta^{-129}$ terms with the spatial resolution of $64^3$ Fourier harmonics.
The MATLAB procedure has been requested to construct the [63/64] Pad\'e
approximants for each entry. The results are shown in \xf{trf}. We observe that
the approximations of the maximum growth rates $\gamma_\alpha$
are relatively accurate for ${\tt tol}=10^{-12}$ and $10^{-16}$
for $\eta\gs0.1$. This bound is roughly 5 times smaller than the minimal $\eta$
for which the power series for $\sAe^l_k$ converge. Table \ref{Tro} sheds light
on the reasons why the gain is unsatisfactory (for $\eta\ge0.1$ spectral
computations of the $\alpha$-effect growth rates for individual $\eta$'s are
efficient): since the rank decreases, when small in absolute value singular values
are discarded, the algorithm ends up with very moderate orders $[2L/2L-1]$.

\begin{table}[t]
\caption{Order parameter $L$ of the Pad\'e approximants $[2L-1/2L]$ (ratios
of polynomials in $1/\eta$) constructed by the algorithm \cite{GGT}
for six independent entries of the symmetrized $\alpha$-effect tensor $\sA$.}
\begin{center}\begin{tabular}{|c|c|c|c|c|c|c|}\hline
{\tt tol}\vphantom{$\displaystyle{\sum}$}
&$\sAe_1^1$&$\sAe_1^2$&$\sAe_1^3$&$\sAe_2^2$&$\sAe_2^3$&$\sAe_3^3$\\\hline
$10^{-10^{\vphantom{|}}}$&5&5&6&5&5&5\\
$10^{-12}$&6&6&6&6&6&6\\
$10^{-14}$&7&7&8&7&7&7\\
$10^{-16}$&8&8&9&8&8&8\\
$10^{-18}$&9&8&10&10&9&9\\
$10^{-20}$&10&10&11&10&10&10\\\hline
\end{tabular}\end{center}\label{Tro}\end{table}

The four remaining panels in \xf{trf} reveal the presence
of the so-called Froissart doublets in the approximants of some entries.
Froissart doublets is a factor of the form $(1/\eta-a)/(1/\eta-\tilde a)$
in the approximant, where the two constants $a$ and $\tilde a$
are close but distinct. Such a factor implies a singular behaviour of the
approximant for $1/\eta$ close to $\tilde a$, not altering much
its behaviour at distances from $\tilde a$ significantly
larger than $|a-\tilde a|$. Often such factors are artifacts emerging due
to noise in the data. Almost vertical segments of the plots are signatures
of the Froissart doublets (see \xf{trf}). They extend to both
positive and negative infinity in the graphs of the approximants, but since
\rf{mgr} are nonlinear functions of the tensor entries, the respective
segments of graphs of $\gamma_\alpha$ may be bounded from below
and/or above. (Their detection has proved unexpectedly difficult; in order
to reliably show their full range in the vertical direction,
we have plotted the approximation step $10^{-9}$ along the abscissa.)
We thus see that although the algorithm \cite{GGT} is supposed to be robust,
it is prone to yield approximants involving Froissart doublets.

\newcounter{foo}
\setcounter{foo}{\value{figure}}

\begin{figure}[p]
\begin{minipage}[t]{.49\textwidth}
\includegraphics[width=\textwidth]{I4512R16.ps}\hspace*{-.4\textwidth}
\raisebox{.25in}{\makebox[.34\textwidth]{\small$L\!=\!4$\hfill(a)}}
\end{minipage}\hfill\begin{minipage}[t]{.49\textwidth}
\includegraphics[width=\textwidth]{I8512R16.ps}\hspace*{-.4\textwidth}
\raisebox{.25in}{\makebox[.34\textwidth]{\small$L\!=\!8$\hfill(b)}}
\end{minipage}

\vspace*{1mm}
\begin{minipage}[t]{.49\textwidth}
\includegraphics[width=\textwidth]{I1464R8.ps}\hspace*{-.4\textwidth}
\raisebox{.25in}{\makebox[.34\textwidth]{\small$L\!=\!14$\hfill(c)}}
\end{minipage}\hfill\begin{minipage}[t]{.49\textwidth}
\includegraphics[width=\textwidth]{I1464R16.ps}\hspace*{-.4\textwidth}
\raisebox{.25in}{\makebox[.34\textwidth]{\small$L\!=\!14$\hfill(d)}}
\end{minipage}

\vspace*{1mm}
\begin{minipage}[t]{.49\textwidth}
\includegraphics[width=\textwidth]{I14512R8.ps}\hspace*{-.4\textwidth}
\raisebox{.25in}{\makebox[.34\textwidth]{\small$L\!=\!14$\hfill(e)}}
\end{minipage}\hfill\begin{minipage}[t]{.49\textwidth}
\includegraphics[width=\textwidth]{I1451216.ps}\hspace*{-.4\textwidth}
\raisebox{.25in}{\makebox[.34\textwidth]{\small$L\!=\!14$\hfill(f)}}
\end{minipage}

\vspace*{1mm}
\begin{minipage}[t]{.49\textwidth}
\includegraphics[width=\textwidth]{I2064R8.ps}\hspace*{-.4\textwidth}
\raisebox{.25in}{\makebox[.34\textwidth]{\small$L\!=\!20$\hfill(g)}}
\end{minipage}\hfill\begin{minipage}[t]{.49\textwidth}
\includegraphics[width=\textwidth]{I2064R16.ps}\hspace*{-.4\textwidth}
\raisebox{.25in}{\makebox[.34\textwidth]{\small$L\!=\!20$\hfill(h)}}
\end{minipage}

\vspace*{1mm}
\begin{minipage}[t]{.49\textwidth}
\includegraphics[width=\textwidth]{I20512R8.ps}\hspace*{-.4\textwidth}
\raisebox{.25in}{\makebox[.34\textwidth]{\small$L\!=\!20$\hfill(i)}}
\end{minipage}\hfill\begin{minipage}[t]{.49\textwidth}
\includegraphics[width=\textwidth]{I2051216.ps}\hspace*{-.4\textwidth}
\raisebox{.25in}{\makebox[.34\textwidth]{\small$L\!=\!20$\hfill(j)}}
\end{minipage}\end{figure}

\setcounter{figure}{\value{foo}}

\begin{figure}[p]
\begin{minipage}[t]{.49\textwidth}
\includegraphics[width=\textwidth]{I2864R8.ps}\hspace*{-.4\textwidth}
\raisebox{.25in}{\makebox[.34\textwidth]{\small$L\!=\!28$\hfill(k)}}
\end{minipage}\hfill\begin{minipage}[t]{.49\textwidth}
\includegraphics[width=\textwidth]{I2864R16.ps}\hspace*{-.4\textwidth}
\raisebox{.25in}{\makebox[.34\textwidth]{\small$L\!=\!28$\hfill(l)}}
\end{minipage}

\vspace*{1mm}
\begin{minipage}[t]{.49\textwidth}
\includegraphics[width=\textwidth]{I28512R8.ps}\hspace*{-.4\textwidth}
\raisebox{.25in}{\makebox[.34\textwidth]{\small$L\!=\!28$\hfill(m)}}
\end{minipage}\hfill\begin{minipage}[t]{.49\textwidth}
\includegraphics[width=\textwidth]{I2851216.ps}\hspace*{-.4\textwidth}
\raisebox{.25in}{\makebox[.34\textwidth]{\small$L\!=\!28$\hfill(n)}}
\end{minipage}

\vspace*{1mm}
\begin{minipage}[t]{.49\textwidth}
\includegraphics[width=\textwidth]{I2964R8.ps}\hspace*{-.4\textwidth}
\raisebox{.25in}{\makebox[.34\textwidth]{\small$L\!=\!29$\hfill(o)}}
\end{minipage}\hfill\begin{minipage}[t]{.49\textwidth}
\includegraphics[width=\textwidth]{I2964R16.ps}\hspace*{-.4\textwidth}
\raisebox{.25in}{\makebox[.34\textwidth]{\small$L\!=\!29$\hfill(p)}}
\end{minipage}

\vspace*{1mm}
\begin{minipage}[t]{.49\textwidth}
\includegraphics[width=\textwidth]{I29512R8.ps}\hspace*{-.4\textwidth}
\raisebox{.25in}{\makebox[.34\textwidth]{\small$L\!=\!29$\hfill(q)}}
\end{minipage}\hfill\begin{minipage}[t]{.49\textwidth}
\includegraphics[width=\textwidth]{I2951216.ps}\hspace*{-.4\textwidth}
\raisebox{.25in}{\makebox[.34\textwidth]{\small$L\!=\!29$\hfill(r)}}
\end{minipage}\end{figure}

\setcounter{figure}{\value{foo}}

\begin{figure}[p]
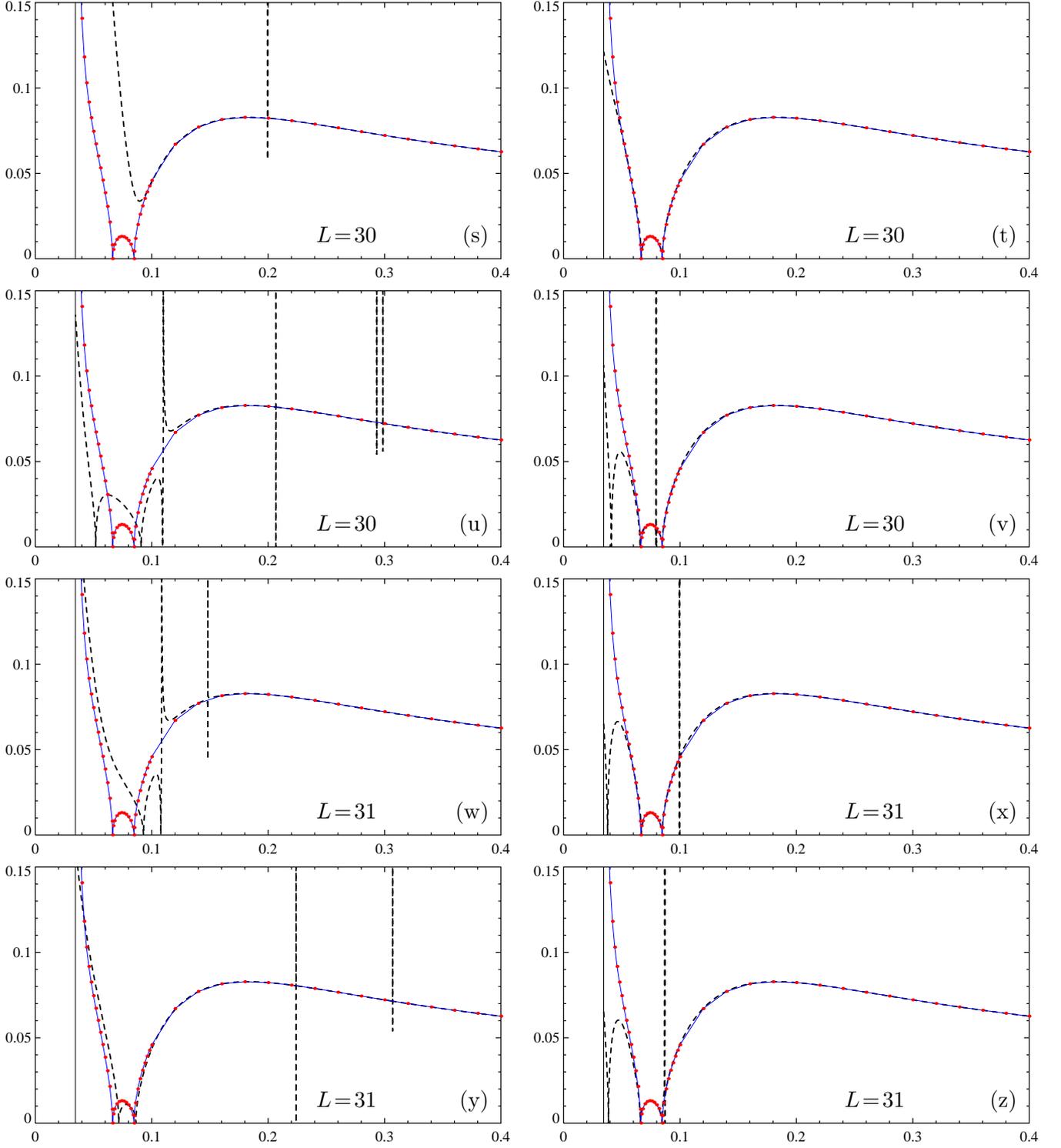

\begin{minipage}[t]{.49\textwidth}
\includegraphics[width=\textwidth]{I3064R8.ps}\hspace*{-.4\textwidth}
\raisebox{.25in}{\makebox[.34\textwidth]{\small$L\!=\!30$\hfill(s)}}
\end{minipage}\hfill\begin{minipage}[t]{.49\textwidth}
\includegraphics[width=\textwidth]{I3064R16.ps}\hspace*{-.4\textwidth}
\raisebox{.25in}{\makebox[.34\textwidth]{\small$L\!=\!30$\hfill(t)}}
\end{minipage}

\vspace*{1mm}
\begin{minipage}[t]{.49\textwidth}
\includegraphics[width=\textwidth]{I30512R8.ps}\hspace*{-.4\textwidth}
\raisebox{.25in}{\makebox[.34\textwidth]{\small$L\!=\!30$\hfill(u)}}
\end{minipage}\hfill\begin{minipage}[t]{.49\textwidth}
\includegraphics[width=\textwidth]{I3051216.ps}\hspace*{-.4\textwidth}
\raisebox{.25in}{\makebox[.34\textwidth]{\small$L\!=\!30$\hfill(v)}}
\end{minipage}

\vspace*{1mm}
\begin{minipage}[t]{.49\textwidth}
\includegraphics[width=\textwidth]{I3164R8.ps}\hspace*{-.4\textwidth}
\raisebox{.25in}{\makebox[.34\textwidth]{\small$L\!=\!31$\hfill(w)}}
\end{minipage}\hfill\begin{minipage}[t]{.49\textwidth}
\includegraphics[width=\textwidth]{I3164R16.ps}\hspace*{-.4\textwidth}
\raisebox{.25in}{\makebox[.34\textwidth]{\small$L\!=\!31$\hfill(x)}}
\end{minipage}

\vspace*{1mm}
\begin{minipage}[t]{.49\textwidth}
\includegraphics[width=\textwidth]{I31512R8.ps}\hspace*{-.4\textwidth}
\raisebox{.25in}{\makebox[.34\textwidth]{\small$L\!=\!31$\hfill(y)}}
\end{minipage}\hfill\begin{minipage}[t]{.49\textwidth}
\includegraphics[width=\textwidth]{I3151216.ps}\hspace*{-.4\textwidth}
\raisebox{.25in}{\makebox[.34\textwidth]{\small$L\!=\!31$\hfill(z)}}
\end{minipage}

\caption{Approximate dependencies of the maximum slow-time growth rates
$\gamma_\alpha$ \rf{mgr} (vertical axis) of large-scale magnetic modes generated
by the $\alpha$-effect on molecular diffusivity $\eta$ (horizontal axis).
Pad\'e approximants $[2L-1/2L]$ of $\alpha$-effect tensor entries are
constructed by the algorithm \cite{nr} for the specified $L$. Resolution $64^3$
(c), (d), (g), (h), (k), (l), (o), (p), (s), (t), (w), (x) and $512^3$ (other
panels) Fourier harmonics, computations with the double (real*8, left panels
except (a)) and quadruple (real*16, right panels and (a)) precision. Thin solid
line: the dependence revealed by computation of $\gamma_\alpha$ at individual
$\eta$ values (red solid circles) by spectral methods (resolution $64^3$
Fourier harmonics), wide dashed line: Pad\'e approximants.}
\label{nrf}\end{figure}

\subsubsection{Approximation by the algorithm \cite{nr}}

Since we have failed to obtain satisfactory resuts with the use
of the algorithm \cite{GGT}, we have also tested the algorithm \cite{nr}.
Quoting from \cite{nr}, although the equations for the coefficients
of the approximant involve a matrix in the Toeplitz form, ``experience shows
that the equations are frequently close to singular, so that one should not
solve them by the methods'' relying on this form, ``but rather by full LU
decomposition. Additionally, it is a good idea to refine the solution
by iterative improvement (routine {\tt mprove} in \S 2.5)''. This
is implemented in their {\tt pade} procedure. We have used it with one
alteration: the routine {\tt mprove} stops when the discrepancy increases;
instead, it has been allowed to make up to 1000 improvement iterations
permitting the discrepancy to temporarily grow and storing the minimum-discrepancy
solution obtained in the course of these iterations (however, it has often been
forced to stop before the allowed number of iterations has been performed,
the iterative process blowing up with an overflow).

Approximate maximum slow-time growth rates \rf{mgr} of large-scale magnetic
modes generated by the $\alpha$-effect have been computed for the same flow
using again $[2L-1/2L]$ Pad\'e approximants for the entries $\sAe_l^k(\eta)$.
They are compared in \xf{nrf} for increasing orders $L$ with the actual maximum
growth rates obtained by direct computation of the fields ${\bf s}_k$ using
spectral methods for individual molecular diffusivities $\eta$.

Four Pad\'e approximants $[2L-1/2L]$ of the entries of the $\alpha$-effect
tensor have been constructed for each considered $L$, using the resolution
of $64^3$ or $512^3$ Fourier harmonics, and running our code with the double
(real*8) or extended quadruple (real*16) precision of the floating-point
number arithmetics. In computers built around Intel and compatible processors,
the former is standard, and the latter is not supported by hardware, but is
software-emulated; however, many compilers do not require modifying the Fortran
source code to use it, all the floating-point data and computations can be
readily promoted to the real*16 precision by using the appropriate compiler
option such as {\tt -r16}. Higher precision and resolution has been expected
to improve the accuracy of the coefficients of the approximants, to augment
the orders of the approximants beyond those produced by the algorithm
\cite{GGT} and to increase the $\eta$
interval, where the growth rate values determined for the approximated
$\alpha$-effect tensor are close to the actual growth rates.

We show in \xf{nrf} the resultant approximations of $\gamma_\alpha$ for $L=4$,
8, 14, 20 and 28 to 31. The four graphs for $L=4$ are visually indistinguishable
and we show only one of them; the same holds true for $L=8$. For $L=14$ and 20,
the plots of the approximated $\gamma_\alpha$, computed with the quadruple
precision for the two spatial resolutions, also visually coincide
(\xf{nrf}(d),(f) and \xf{nrf}(h),(j)), but this is wrong
for the respective double precision approximations. For higher $L$, all four
plots are visually distinct. The quadruple precision approximations are
plagued much less by the Froissart doublets (and never involve multiple
occurrences of the doublets) than the double precision ones.
Three high-$L$ quadruple precision approximations are
reasonably accurate: for $L=28$ and the $512^3$ resolution for $\eta\ge0.055$
(\xf{nrf}(n)); for $L=29$ and the $64^3$ resolution for $\eta\ge0.05$ (\xf{nrf}(p));
and for $L=30$ and the $64^3$ resolution for $\eta\ge0.05$ (\xf{nrf}(t)). Thus,
the left end of the interval of validity of Pad\'e approximations has decreased
roughly twice compared to that obtained by the algorithm \cite{GGT}. All other
quadruple precision approximations for $L\ge29$ (\xf{nrf}(r), (v), (x) and (z))
can also give reasonable accuracy for $\eta\gs0.06\div0.07$ upon removal
of Froissart doublets from the affected approximants of $\sAe_l^k$.

These results suggest, that Pad\'e approximants are useful for representing
the functional dependence of the slow-time growth rates \rf{mgr} of large-scale
magnetic modes generated by the $\alpha$-effect for fairly low magnetic
molecular diffusivities. However, for construction of Pad\'e approximants,
accurate enough to serve small $\eta$, the quadruple precision arithmetics
must be used, and hence run times become comparable to those of direct
computation of the growth rates at individual $\eta$'s (note that real*16
computations are typically ten times slower than real*8 ones). Therefore another
strategy is perhaps also sensible: in computations for an individual $\eta$,
to use relatively low-order not-very-precise Pad\'e approximants for neutral
modes ${\bf s}_k$ (constructed, for instance, at each point in space or
for each Fourier harmonics within the employed resolution) as
the initial data for further refinement by the usual iterative methods.

\section{Computation of the magnetic eddy diffusivity tensor}\label{sec:med}

As discussed in section \ref{sec:edvis},
an important class are parity-invariant flows. This
symmetry is compatible with the equations of fluid dynamics (the Navier--Stokes
or Euler equations) provided the forcing has the same property. For such a flow,
the domain of the operator of magnetic induction $\L$ splits into the subspaces
of parity-invariant fields (such that ${\bf f}(-{\bf x})=-{\bf f}({\bf x})$)
and of parity-antiinvariant ones (such that ${\bf f}(-{\bf x})={\bf f}({\bf x})$).
Solutions to the auxiliary problems \rf{Seq}, ${\bf s}_k({\bf x})$,
are therefore parity-antiinvariant. This implies $\A=0$ (see \rf{Adef}),
i.e., no $\alpha$-effect acts in such flows, and hence $\lambda_1=0$.

\subsection{The multiscale formalism revealing the magnetic eddy diffusivity}

By \rf{eq1},
$${\bf b}_1=\sum_{k=1}^3\sum_{m=1}^3{\partial\LA{\bf b}_0\RA_k\over\partial X_m}
\,{\bf g}_{mk},$$
where the small-scale zero-mean (non-solenoidal!) fields
${\bf g}_{mk}({\bf x})$ solve nine {\it auxiliary problems of type~II}:
\BE\L{\bf g}_{mk}=-2\eta{\partial{\bf s}_k\over\partial x_m}
-{\bf e}_m\times({\bf v}\times{\bf s}_k).\EE{Geq}
For parity-invariant $\bf v$, ${\bf g}_{mk}({\bf x})$ are also parity-invariant;
moreover, ${\bf b}_n$ are parity-antiinvariant for all even $n$ and
parity-invariant for odd $n$ \cite{VZ} in the expansion \rf{bex}, and no odd
powers of $\varepsilon$ enter the series \rf{lex} for the eigenvalue $\lambda$.

Averaging the third (order $\varepsilon^2$) equation in the hierarchy yields
\BE\eta\nabla^2_{\bf X}\LA{\bf b}_0\RA+\nabla_{\bf X}\times\sum_{k=1}^3\sum_{m=1}^3\D_{mk}
{\partial\LA{\bf b}_0\RA_k\over\partial X_m}=\lambda_2\LA{\bf b}_0\RA,\EE{med}
where
\BE\D_{mk}=\LA{\bf v}\times{\bf g}_{mk}\RA\EE{Ddef}
is the so-called tensor of magnetic eddy diffusivity correction. Again assuming
that the mean field $\LA{\bf b}_0\RA$ is a Fourier harmonics \rf{mm}, we find
\cite{RCZ}
\se{dei}\be\lambda_{2_\pm}({\bf q})&=-\eta-{1\over2}\sum_{j,l,n}
(D^l_n-D^n_l)q_j\pm\sqrt d,\label{d2}\\
d&=\sum_{j,l,n}\left(((\sD^l_n)^2-\sD^l_l\,\sD^n_n)q_j^2-2q_jq_n(\sD^l_n\,
\sD^l_j-\sD^l_l\,\sD^n_j)\right),\label{d3}\end{align}
where both sums are over even permutations of indices 1, 2 and 3 (i.e.,
$(j,l,n)$ are combinations (1,2,3), (2,3,1) and (3,1,2)) and it is denoted
\BE D^l_n=\sum_m\De^l_{mn}q_m,\qquad\sD^l_n=(D^l_n+D^n_l)/2.\EE{d1}\end{subequations}
The minimum
\BE\eta_{\rm eddy}\equiv\min_{|{\bf q}|=1}(-{\rm Re}\,\lambda_{2_\pm}({\bf q}))\EE{miE}
is called the minimum magnetic eddy diffusivity; when it is negative,
the interaction of the fluctuating small-scale velocity and magnetic field is
capable of generating large-scale magnetic fields. For this reason, this
quantity is of prime interest in the large-scale dynamo theory.

Expression \rf{Ddef} can be transformed into
\BE\De^l_{mk}=\LA{\bf Z}_l\cdot\left(2\eta{\partial{\bf s}_k\over\partial x_m}
+{\bf e}_m\times({\bf v}\times{\bf s}_k)\right)\RA,\EE{Dlmk}
where ${\bf Z}_l$ are zero-mean solutions
to three {\it auxiliary problems for the adjoint operator}:
\BE\L^*{\bf Z}_l={\bf v}\times{\bf e}_l\EE{adj}
and $\L^*:{\bf z}\mapsto\eta\nabla^2{\bf z}-\{{\bf v}\times(\nabla\times{\bf z})\}$
is the operator adjoint to $\L$ acting in the space of zero-mean
space-periodic fields.

\subsection{Pad\'e approximation}

Relation \rf{Ddef} suggests to construct expansions of the solutions
to the auxiliary problems in the inverse molecular diffusivity, \rf{Se} and
\BE{\bf g}_{mk}=\sum^\infty_{n=1}{\bf g}_{mk}^{(n)}\eta^{-n}.\EE{Ge}
Dividing \rf{Geq} by $\eta$ yields
$$\nabla^2{\bf g}_{mk}=-\eta^{-1}\nabla\times({\bf v}\times{\bf g}_{mk})
-2{\partial{\bf s}_k\over\partial x_m}
-\eta^{-1}{\bf e}_m\times({\bf v}\times{\bf s}_k),$$
whereby
\se{rrG}\be{\bf g}_{mk}^{(1)}&=-\nabla^{-2}\left(2\,{\partial{\bf s}_k^{(1)}
\over\partial x_m}+{\bf e}_m\times({\bf v}\times{\bf e}_k)\right),\label{G1}\\
{\bf g}_{mk}^{(n)}&=-\nabla^{-2}\left(\nabla\times({\bf v}\times
{\bf g}_{mk}^{(n-1)})+2\,{\partial{\bf s}_k^{(n)}\over\partial x_m}
+{\bf e}_m\times({\bf v}\times{\bf s}_k^{(n-1)})\right)\quad\mbox{for~}n>1.
\label{Gn}\end{align}\end{subequations}
Clearly, the flow being parity-invariant, all ${\bf s}_k^{(n)}$ are
parity-antiinvariant and all ${\bf g}_{mk}^{(n)}$ parity-invariant.
By \rf{Ddef} and \rf{Ge},
\BE\D_{mk}=\sum^\infty_{n=1}\D_{mk}^{(n)}\eta^{-n},\qquad
\D_{mk}^{(n)}=\LA{\bf v}\times{\bf g}_{mk}^{(n)}\RA.\EE{De}
Thus, algorithm I consists of the following steps:\\
$\bullet$ find the fields ${\bf s}_k^{(n)}$ employing \rf{rrS};\\
$\bullet$ find the fields ${\bf g}_{mk}^{(n)}$ employing \rf{rrG};\\
$\bullet$ calculate the coefficients $\D_{mk}^{(n)}$ employing \rf{De}.

Expressions \rf{De} reveal symmetry properties of coefficients in the eddy
diffusivity tensor expansion (similar to those of the coefficients of the
$\alpha$-effect tensor expansion). Eddy diffusivity tensor for the reverse
flow $-\bf v$ is related to that of the flow $\bf v$ by the relations
${(\D^-_{ml})}_k=-\De_{mk}^l$ \cite{ABNZ}. By \rf{rrG},
${\bf g}_{mk}^{(n)}=(-1)^n({\bf g}_{mk}^-)^{(n)}$. Thus, identities, analogous
to those used in the case of the $\alpha$-effect tensor, reveal that
for each fixed $m$ the coefficients $(\De_{mk}^l)^{(n)}$ in the series \rf{De}
are symmetric $3\times3$ matrices for even $n$, and antisymmetric
ones for odd $n$. This implies that the symmetrized matrix $\mD$ \rf{d1},
determining the discriminant $d$ \rf{d3}, is expanded in even
powers of $1/\eta$, and the antisymmetric one $\bf D-\mD$, determining
the common part of $\lambda_{2_\pm}({\bf q})$ \rf{d2}, in odd powers.

It is simple to show that, like in the case of $\alpha$-effect
dynamos, for a given flow $\bf v$ the radius of convergence of all the series
\rf{Ge} and \rf{De} (regarded as functions of $1/\eta$) is generically equal to
$1/\max_i|\mu_i|$, where $\mu_i$ are eigenvalues of the compact operator $\M$
\rf{rr}; convergence of the series is guaranteed for $\eta>\max_i|\mu_i|$.

An alternative form of the eddy diffusivity tensor can be exploited.
Comparing \rf{Seq} and~\rf{adj}, we find
\BE{\bf e}_l+\nabla\times{\bf Z}_l={\bf s}^-_l\quad\Rightarrow\quad
{\bf Z}_l=\eta^{-1}\nabla^{-2}({\bf v}\times{\bf s}^-_l),\EE{Zl}
where the superscript minus denotes objects pertinent to the reverse flow
$-\bf v$ (see \rf{iv}). Using \rf{Zl} to eliminate ${\bf s}_k$ in \rf{Dlmk}
yields \cite{ABNZ}
\BE\De^l_{mk}=\eta\LA{\bf Z}_l\cdot\left(2\,\nabla\times{\partial{\bf Z}^-_k
\over\partial x_m}-{\bf e}_m\times\nabla^2{\bf Z}^-_k\right)\RA.\EE{ZlZk}
By \rf{Zl}, all ${\bf Z}_l$ are parity-invariant, and hence \rf{adj} is
equivalent to
$$\nabla^2{\bf Z}_l=\eta^{-1}({\bf v}\times({\bf e}_l+\nabla\times{\bf Z}_l)),$$
which implies a power series expansion
\BE{\bf Z}_l=\sum^\infty_{n=1}{\bf Z}_l^{(n)}\eta^{-n},\EE{Ze}
where the coefficients satisfy recurrence relations
\BE{\bf Z}_l^{(1)}=\nabla^{-2}({\bf v}\times{\bf e}_l),\qquad
{\bf Z}^{(n)}_l=\nabla^{-2}({\bf v}\times(\nabla\times{\bf Z}^{(n-1)}_l))
\quad\mbox{for~}n>1.\EE{rrZ}
By linearity of \rf{rrZ} in $\bf v$,
the coefficients of such an expansion for the reverse flow are linked:
$$({\bf Z}^-_l)^{(n)}=(-1)^n{\bf Z}^{(n)}_l.$$
This implies algorithm II for calculation of $(\De^l_{mk})^{(n)}$
based on \rf{ZlZk}:\\
$\bullet$ find the fields ${\bf Z}_l^{(n)}$ applying \rf{rrZ};\\
$\bullet$ for $n\ge1$ calculate
\BE(\De^l_{mk})^{(n)}=\sum_{j=1}^n(-1)^j\LA{\bf Z}^{(n+1-j)}_l\cdot
\left(2\,\nabla\times{\partial{\bf Z}^{(j)}_k\over\partial x_m}
-{\bf e}_m\times\nabla^2{\bf Z}^{(j)}_k\right)\RA.\EE{jZ}

Using \rf{Zl}, it is straightforward albeit tedious to transform \rf{ZlZk} into
\BE\De^l_{mk}=\eta\LA\{{\bf s}_l^-\}\times\{{\bf s}_k\}-\{{\bf s}_k\}\nabla
\cdot{\bf Z}_l+\{{\bf s}^-_l\}\nabla\cdot{\bf Z}^-_k\RA_m.\EE{Dn}
Relations \rf{Zl} imply
\BE\nabla\times{\bf Z}_l^{(n)}=(-1)^n{\bf s}^{(n)}_l,\qquad{\bf Z}^{(n)}_l=
(-1)^{n-1}\nabla^{-2}({\bf v}\times{\bf s}^{(n-1)}_l)\quad\mbox{for~}n\ge1.\EE{Zc}
Thus the coefficients in the expansion \rf{De} have the entries
\BE(\De^l_{mk})^{(n)}=\sum_{j=1}^n\LA(-1)^j{\bf s}_l^{(j)}\times
{\bf s}_k^{(n+1-j)}-{\bf s}^{(n+1-j)}_k\nabla\cdot{\bf Z}^{(j)}_l
-(-1)^n{\bf s}^{(n+1-j)}_l\nabla\cdot{\bf Z}^{(j)}_k\RA_m.\EE{Dc}
Algorithm III consists of the following steps:\\
$\bullet$ determine coefficients in the expansion of the neutral magnetic
modes ${\bf s}_k$ applying recurrence relations \rf{rrS};\\
$\bullet$ in the course of these calculations, determine coefficients
in the expansion of $\nabla\cdot{\bf Z}_l$ using \rf{Zc};\\
$\bullet$ calculate the coefficients $(\De^l_{mk})^{(n)}$ applying \rf{Dc}.

If the flow $\bf v$ involves a small number of Fourier harmonics, it is unclear
a priori, which of the three algorithms is more efficient. Algorithm I involves
computation of two sets of coefficients, for ${\bf s}_k$ and ${\bf g}_{mk}$;
algorithm II only of the set of coefficients for ${\bf Z}_l$; algorithm III is
an intermediate case involving two sets of coefficients, for ${\bf s}_k$ and
$\nabla\cdot{\bf Z}_l$. However, calculation of the $n$-term sums \rf{jZ} and
\rf{Dc} in algorithms II and III, respectively, is computer-intensive.

\subsection{Numerical results}

\begin{figure}[t]
\centerline{\includegraphics[height=65mm]{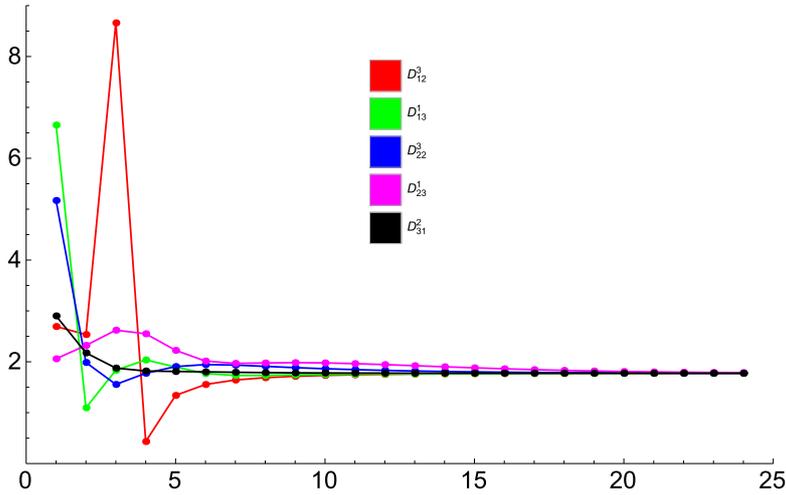}}
\caption{The ratios $|(\De^l_{mk})^{(2n-1)}/(\De^l_{mk})^{(2n+1)}|^{1/2}$
(vertical axis) versus $n>0$ (horizontal axis) for five independent entries
of the eddy diffusivity tensor for the cosine flow \rf{cos}, \rf{coe}.}
\label{rad}\end{figure}

For numerical experimentation we have applied two types of flows:
the so-called {\it cosine flows} introduced in \cite{RCZ}, and their curls
considered in \cite{ACZ2}. They are of interest in that the latter have
a pointwise zero vorticity (kinematic) helicity, and the former have
a pointwise zero velocity helicity, and nevertheless they are capable of both
small- and large-scale magnetic field generation (see {\it ibid}). Involving
a small number of trigonometric functions, they are particularly useful
for calculating the Taylor series coefficients \rf{De}.

The cosine flows are defined as follows:
\be v_1&=\beta n(b_1\sin({\bf a\cdot x})+a_1\sin({\bf b\cdot x}))\cos nx_3,\nonumber\\
v_2&=\beta n(b_2\sin({\bf a\cdot x})+a_2\sin({\bf b\cdot x}))\cos nx_3,\label{cos}\\
v_3&=-\beta({\bf a\cdot b})(\cos({\bf a\cdot x})+\cos({\bf b\cdot x}))\sin nx_3.
\nonumber\end{align}
Here ${\bf a}=(a_1,a_2,0)$ and ${\bf b}=(b_1,b_2,0)$ are constant horizontal
vectors, and
$$\beta=2(n^2(|{\bf a}|^2+|{\bf b}|^2)+2({\bf a\cdot b})^2)^{-1/2},$$
so that the r.m.s.~flow velocity is unity.

\begin{figure}[t]
\begin{minipage}[t]{.49\textwidth}
\includegraphics[width=\textwidth]{Pad2020q.ps}\hspace*{-.4\textwidth}
\raisebox{.25in}{\makebox[.34\textwidth]{\hfill(a)}}
\end{minipage}\hfill\begin{minipage}[t]{.49\textwidth}
\includegraphics[width=\textwidth]{q1q2q3.ps}\hspace*{-.4\textwidth}
\raisebox{.25in}{\makebox[.34\textwidth]{\hfill(b)}}
\end{minipage}

\vspace*{1mm}
\begin{minipage}[t]{.49\textwidth}
\includegraphics[width=\textwidth]{Par20-18.ps}\hspace*{-.4\textwidth}
\raisebox{.25in}{\makebox[.34\textwidth]{\hfill(c)}}
\end{minipage}\hfill\begin{minipage}[t]{.49\textwidth}
\includegraphics[width=\textwidth]{Pad2424q.ps}\hspace*{-.4\textwidth}
\raisebox{.25in}{\makebox[.34\textwidth]{\hfill(d)}}
\end{minipage}
\caption{Minimum eddy diffusivity \rf{edq} (vertical axis) for the sample
flow \rf{cos}, \rf{coe} computed using Pad\'e approximants of the quantities $q_i$
\rf{qq} of orders [20/20] (a), [18/18] (green line) and regularized [20/20]
(black line) (c), and [24/24]~(d). Behaviour of the [20/20] Pad\'e approximants
of $q_i$ (vertical axis) near the points, where Froissart doublets
are located (b). Horizontal axis: magnetic molecular diffusivity $\eta$.
Red dots: minimum eddy diffusivity computed by spectral methods (resolution
$128^3$ Fourier harmonics) individually for the respective $\eta$ values.}
\label{paq}

~

\centerline{\includegraphics[width=.5\textwidth]{Pad2322D.ps}}
\caption{Minimum eddy diffusivity \rf{edq} (vertical axis) for the sample
flow \rf{cos}, \rf{coe} computed by Pad\'e approximation of the individual
entries of the eddy diffusivity tensor $\D$ of orders [23/22] for varying
magnetic molecular diffusivity $\eta$ (horizontal axis).}
\label{paD}\end{figure}

Because of many symmetries of the cosine flows, all entries of the eddy
diffusivity tensor $\D$ vanish, except for five pairs (see \cite{RCZ}):
$$\De^2_{31}=-\De^1_{32},~~\De^3_{12}=-\De^2_{13},~~
\De^1_{23}=-\De^3_{21},~~\De^3_{22}=-\De^2_{23},~~
\De^1_{13}=-\De^3_{11}.$$
Consequently, the minimum eddy diffusivity \rf{miE} takes a simple form:
\BE\eta_{\rm eddy}=\eta-\max\left(\De^2_{31},\ {1\over2}\left(\De^3_{12}+\De^1_{23}
+\sqrt{(\De^3_{12}-\De^1_{23})^2+(\De^3_{22}+\De^1_{13})^2}\right)\right).\EE{Sed}

We have considered the particular sample flow for
\BE{\bf a}=(1,0,0),\qquad{\bf b}=(1,1,0),\qquad n=1\EE{coe}
and used Mathematica again to implement algorithm I: we have calculated exactly
the coefficients of the expansions \rf{Se} and \rf{Ge} of solutions
to the auxiliary problems \rf{Seq} and \rf{Geq} using the recurrence relations
\rf{rrS} and \rf{rrG}, and of the coefficients $\D_{mk}^{(n)}$ of the series
\rf{De} up to order~$\eta^{-49}$. The~precise coefficients ${\bf s}_k^{(49)}$
and ${\bf g}_{mk}^{(49)}$ require
about 2~Gbytes of memory for storage (in the ASCII form). For the flow \rf{cos},
the coefficients $\D_{mk}^{(n)}$ turn out to be rational; in the ASCII form,
the vectors $\D_{mk}^{(49)}$ occupy 10 to 20 Kbyte of memory.

\xf{rad} shows the sequence of the ratios
$|(\De^l_{mk})^{(2n-1)}/(\De^l_{mk})^{(2n+1)}|^{1/2}$ for five independent
entries. The limit of this sequence for $n\to\infty$ is equal to the radius
of convergence of the series \rf{De} (regarded as a function of $1/\eta$)
for the respective entry (note that due to the antisymmetry in $l$ and $k$,
the entries involve only
odd powers of $1/\eta$). The figure demonstrates that the series for the five
entries have the same radius of convergence and converge for $\eta\gs0.5$.

Because precise Mathematica calculations require considerable computer
resources, we have not considered high-order Pad\'e approximants.
The amount of calculations reduces if meromorphic functions
\BE q_1=\De^2_{31},\qquad q_2=(\De^3_{12}+\De^1_{23})/2,\qquad
q_3=((\De^3_{12}-\De^1_{23})^2+(\De^3_{22}+\De^1_{13})^2)/4,\EE{qq}
are approximated only, in terms of which (see \rf{Sed})
\BE\eta_{\rm eddy}=\eta-\max(q_1,q_2+\sqrt{q_3}).\EE{edq}
The poor quality of the resultant approximation (see \xf{paq}(a)) is due to
the presence of two Froissart doublets in the approximants of $q_2$ and $q_3$
(\xf{paq}(b)). Gaps are present in the plot in \xf{paq}(a), where
the approximant of $q_3$ becomes negative due to its singular behaviour and
thus the square root in \rf{edq} can not be extracted. Upon factoring
the doublets out (which is simple in Mathematica) in the two plagued
approximants, the quality of the approximation becomes very similar to that
obtained by using [18/18] approximants of $q_i$ (see \xf{paq}(c)). Increasing
the orders to [24/24] does not significantly improve the approximated
$\eta_{\rm eddy}$ (see \xf{paq}(d)). A better approximation is obtained if
the elements of the eddy diffusivity tensor $\D$ are Pad\'e-approximated
individually (see \xf{paD}); this gives reasonably accurate values of
$\eta_{\rm eddy}$ for $\eta\gs0.02$, which is roughly 25 times larger than
the minimum $\eta$, for which the power series in $1/\eta$ for the fields
${\bf s}_k$ and ${\bf g}_{mk}$, as well as for the elements of the tensor $\D$
are convergent.

\begin{figure}[t]
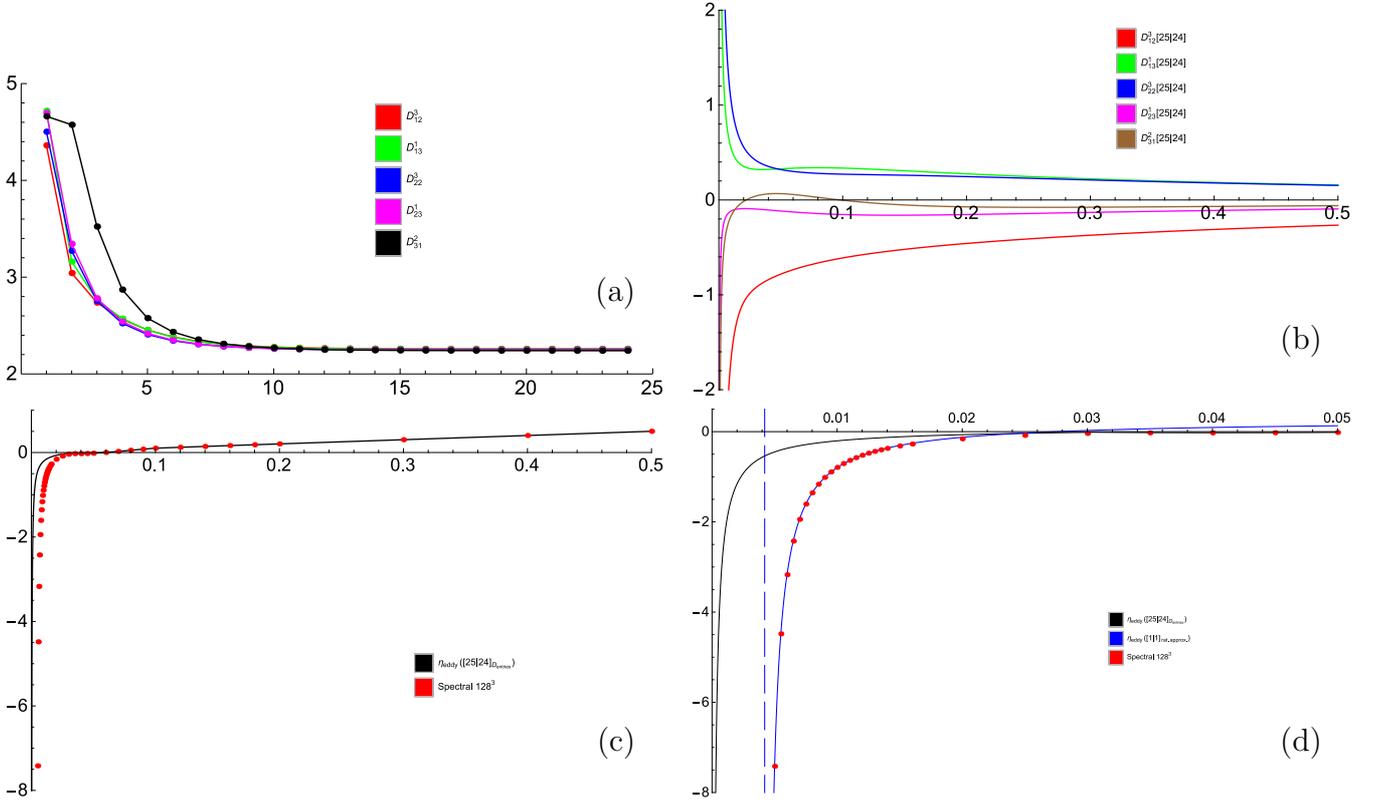

\begin{minipage}[t]{.49\textwidth}
\includegraphics[width=\textwidth]{radius2.ps}\hspace*{-.4\textwidth}
\raisebox{.5in}{\makebox[.34\textwidth]{\hfill(a)}}
\end{minipage}\hfill\begin{minipage}[t]{.49\textwidth}
\includegraphics[width=\textwidth]{DSnew.ps}\hspace*{-.4\textwidth}
\raisebox{.25in}{\makebox[.34\textwidth]{\hfill(b)}}
\end{minipage}

\vspace*{1mm}
\begin{minipage}[t]{.49\textwidth}
\includegraphics[width=\textwidth]{EtaEddy.ps}\hspace*{-.4\textwidth}
\raisebox{.25in}{\makebox[.34\textwidth]{\hfill(c)}}
\end{minipage}\hfill\begin{minipage}[t]{.49\textwidth}
\includegraphics[width=\textwidth]{hyperb.ps}\hspace*{-.4\textwidth}
\raisebox{.25in}{\makebox[.34\textwidth]{\hfill(d)}}
\end{minipage}
\caption{The ratios $|(\De^l_{mk})^{(2n-1)}/(\De^l_{mk})^{(2n+1)}|^{1/2}$
(vertical axis) versus $n>0$ (horizontal axis) for five independent entries
of the eddy diffusivity tensor for the flow \rf{c2}, \rf{co2} (a).
The [25/24] Pad\'e approximants for the five entries $\De^l_{mk}$
(vertical axis) versus $\eta$ (horizontal axis) (b) and minimum eddy
diffusivity \rf{edq} (vertical axis) computed using these approximants
for varying $\eta$ (horizontal axis) (c).
Red dots: minimum eddy diffusivity computed by spectral methods (resolution
$128^3$ Fourier harmonics) individually for the respective $\eta$ values.
Zoom of the plot (c) for small $\eta$ (black line) and a hyperbolic fit
(blue line) through the 20 spectral eddy diffusivity values for 20
$\eta$ points in the interval $0.0045\le\eta\le0.014$ step 0.0005 (d).
The dashed line shows the vertical asymptote of the minimum eddy diffusivity
at the onset of small-scale dynamo in the symmetry subspace, where the
neutral mode ${\bf s}_3$ resides.}
\label{pa2}\end{figure}

Upon shifting by a quarter of the period in the vertical coordinate $x_3$,
the curl of \rf{cos} takes the form
\be v_1=&\,\beta\left((({\bf a\cdot b})a_2+n^2b_2)\sin({\bf a\cdot x})
+(({\bf a\cdot b})b_2+n^2a_2)\sin({\bf b\cdot x})\right)\cos nx_3,\nonumber\\
v_2=&-\beta\left((({\bf a\cdot b})a_1+n^2b_1)\sin({\bf a\cdot x})
+(({\bf a\cdot b})b_1+n^2a_1)\sin({\bf b\cdot x})\right)\cos nx_3,\label{c2}\\
v_3=&\,\beta n(a_2b_1-a_1b_2)(\cos({\bf a\cdot x})-\cos({\bf b\cdot x}))
\sin nx_3,\nonumber\end{align}
where we now assume the normalizing factor
$$\beta=2\left((n^4+({\bf a\cdot b})^2)(|{\bf a}|^2+|{\bf b}|^2)
+2n^2(({\bf a\cdot b})^2+|{\bf a}|^2|{\bf b}|^2)\right)^{-1/2},$$
for which the r.m.s.~flow velocity is again 1. Since this flow possesses
all the symmetries of \rf{cos}, the expression \rf{Sed} for the minimum eddy
diffusivity still applies.

Following algorithm II, for a sample flow \rf{c2} for
\BE{\bf a}=(0,1,0),\qquad{\bf b}=(2,2,0),\qquad n=1,\EE{co2}
we have calculated by Mathematica 49 coefficients $\D_{mk}^{(n)}$ of the series
\rf{De} up to order~$\eta^{-49}$.
The graph of the ratios $|(\De^l_{mk})^{(2n-1)}/(\De^l_{mk})^{(2n+1)}|^{1/2}$
for the five independent entries (see \xf{pa2}(a)) of the eddy diffusivity
tensor shows that the series \rf{De} converge for $\eta\gs2.2$.
The highest-order (for this set of coefficients) [25/24] approximants
of the entries are free of Froissart doublets (see \xf{pa2}(b)). They yield
a satisfactory approximation of dependence on $\eta$ of the minimum magnetic
eddy diffusivity for $\eta\gs0.03$, which is roughly 70 times smaller
than the bound obtained for convergence of the Taylor series \rf{De}
for $\De^l_{mk}$ (see \xf{pa2}(c),(d)). However, their fidelity
is insufficient to reproduce the singularity of the minimum eddy diffusivity
observed in \xf{pa2}(c),(d).

The symmetries of the flow \rf{c2}, \rf{co2} imply, that the neutral modes
${\bf s}_k$ reside in invariant subspaces of the magnetic induction operator,
that can be categorized in terms of the Fourier harmonics
${\bf b_n}\e^{\I\bf n\cdot x}$, comprising the Fourier series for ${\bf s}_k$
(by virtue of the mode periodicity, the wave vectors $\bf n$ have integer
components). Only the harmonics that have the following properties enter
the Fourier series for ${\bf s}_k$:\\
$\bullet$ $\bf b_n=b_{-n}$ are real; \\
$\bullet$ the numbers $n_1$ and $n_1/2+n_2+n_3$ are even;\\
$\bullet$ ${\bf s}_1$ and ${\bf s}_2$ are symmetric in $x_3$ (i.e.,
$b_{\bf n}^1=b_{{\bf n}^*}^1,~b_{\bf n}^2=b_{{\bf n}^*}^2,~
b_{\bf n}^3=-b_{{\bf n}^*}^3$) and ${\bf s}_3$ is antisymmetric in $x_3$
(i.e., $b_{\bf n}^1=-b_{{\bf n}^*}^1,~b_{\bf n}^2=-b_{{\bf n}^*}^2,~
b_{\bf n}^3=b_{{\bf n}^*}^3$), where ${\bf n}^*=(n_1,n_2,-n_3)$.\\
We have checked that for $\eta>\eta_{\rm cr}=0.00420516$ small-scale modes
from the subspace, where ${\bf s}_1$ and ${\bf s}_2$ are located, are not
generated, but at $\eta=\eta_{\rm cr}$ the small-scale generation starts
in the subspace, where ${\bf s}_3$ resides. It is known \cite{ZPF,RCZ,ACZ2}
that the point of the onset of the small-scale generation is typically
associated with a singularity of the $\alpha$-effect or eddy diffusivity
tensors; this is the case for the flow under consideration. We observe
in \xf{pa2}(d) that the least-squares fit by a hyperbola through 20 computed
values of minimum eddy diffusivity at equispaced points in the interval
$0.0045\le\eta\le0.014$ is very accurate (actually, only 19 points out of
the 20 are shown; for the smallest $\eta=0.0045$, eddy diffusivity
-20.217501, also well approximated by the hyperbola, is out of the vertical
range of \xf{pa2}(d)). The hyperbolic fit yields the location of the singularity
at $\eta=0.0041988$ (the vertical asymptote is shown by a dashed line
in \xf{pa2}(d)) which is very close to the point of the onset
of the small-scale generation $\eta_{\rm cr}=0.00420516$ computed by spectral
methods; the hyperbola through the three smallest $\eta$ from this interval
yields a closer value 0.00420233 .

\section{Conclusions}\label{sec:co}

We have tested Pad\'e approximants of the $\alpha$-effect and eddy diffusivity
tensors, responsible for generation of large-scale fields, as functions of
the respective molecular diffusivity: the viscosity~$\nu$
when hydrodynamic perturbations are studied, and the magnetic diffusivity
$\eta$ when kinematic dynamo problem is under scrutiny. We have tried
different computational tools: Fortran codes relying on the
floating point arithmetics, and Mathematica for symbolic and arbitrary
precision calculations. A relatively high (several dozens) order of Pad\'e
approximants is needed to obtain a reasonable accuracy of approximation
of the tensor entries. For this, high precision of computations
(in particular, the quadruple precision in Fortran) has proved indispensable.
For our sample flows the Pad\'e-approximated tensors yield
large-scale magnetic field growth rates to satisfactory accuracy
for $\eta$, several dozen times smaller then those, for which
power series in the inverse molecular diffusivity converge, for both
large-scale generating mechanisms (the $\alpha$-effect
and negative magnetic eddy diffusivity).

Application of these techniques in computational fluid dynamics
and magnetohydrodynamics seems natural for estimating transport coefficients
quantifying the influence of small scales on the evolution of large-scale
fields in the spirit of Large Eddy Simulation methods. Our findings, while
promising, suggest that to achieve this goal additional algorithms
are needed for determination\\
$\bullet$ of Froissart doublets in approximants of tensor entries
and their elimination (the approach of \cite{BLM} may prove useful
for monitoring the absence of the doublets);\\
$\bullet$ of the interval in molecular diffusivity, where
the approximation is sufficiently accurate;\\
$\bullet$ of the realistic orders of a Pad\'e approximant, for which the length
of such interval is close to the maximum.\\
It is relatively easy to perform these tasks manually by trial-and-error
methods --- the difficulty lies in performing them automatically.

\section*{Acknowledgements}
This work was supported by CMUP (Centro de Matem\'atica da
Universidade do Porto, UID/ MAT/00144/2019) (SG$+$VZ) and SYSTEC (Centro
de Investiga\c c\~ao em Sistemas e Tecnologias,
POCI-01-0145-FEDER-006933/SYSTEC) (RC), which are funded
by FCT with national (MCTES) and European structural funds through the programs
FEDER (Fundo Europeu de Desenvolvimento Regional / European Regional
Development Fund) under the partnership agreement PT2020, and projects STRIDE
[NORTE-01-0145-FEDER-000033]
funded by FEDER -- NORTE 2020 (SG$+$VZ $+$RC) and MAGIC
[POCI-01-0145-FEDER-032485] funded by FEDER via COMPETE 2020 -- POCI (SG$+$VZ).

\end{document}